\documentclass[twocolumn]{aastex61}

\usepackage{color}
\usepackage{graphicx}
\usepackage{subfigure}
\usepackage{topcapt}
\usepackage{float}
\usepackage{natbib}
\usepackage{amsmath}
\usepackage{url}
\usepackage{amssymb}
\usepackage{lineno}
\graphicspath{ {figs/} }

\newcommand{\nustar} {\textit{NuSTAR}}
\newcommand{\foxsi} {\textit{FOXSI}}
\newcommand{\rhessi} {\textit{RHESSI}}

\newcommand{\angstrom}{\mbox{\normalfont\AA}}

\begin{document}

\title{Hard X-Ray Constraints on Small-Scale Coronal Heating Events}

\correspondingauthor{David M. Smith}
\email{dsmith8@ucsc.edu}
\author{Andrew J. Marsh}
\affiliation{NextEra Energy Resources, Juno Beach, FL 33408, USA}
\affiliation{Santa Cruz Institute for Particle Physics and Department of Physics, University of California, Santa Cruz, CA 95064, USA}
\author{David M. Smith}
\affiliation{Santa Cruz Institute for Particle Physics and Department of Physics, University of California, Santa Cruz, CA 95064, USA}
\author{Lindsay Glesener}
\affiliation{School of Physics \& Astronomy, University of Minnesota Twin Cities, Minneapolis, MN 55455, USA}
\author{James A. Klimchuk}
\affiliation{NASA Goddard Space Flight Center, Solar Physics Lab., Greenbelt, MD 20771, USA}
\author{Stephen J. Bradshaw}
\affiliation{Department of Physics and Astronomy, Rice University, Houston, TX 77005, USA}
\author{Juliana Vievering}
\affiliation{School of Physics \& Astronomy, University of Minnesota Twin Cities, Minneapolis, MN 55455, USA}
\author{Iain G. Hannah}
\affiliation{SUPA School of Physics \& Astronomy, University of Glasgow, Glasgow G12 8QQ, UK}
\author{Steven Christe}
\affiliation{NASA Goddard Space Flight Center, Solar Physics Lab., Greenbelt, MD 20771, USA}
\author{Shin-nosuke Ishikawa}
\affiliation{Institute of Space and Astronautical Science, Japan Aerospace Exploration Agency}
\author{S\"{a}m Krucker}
\affiliation{Space Sciences Laboratory University of California, Berkeley, CA 94720, USA}
\affiliation{University of Applied Sciences and Arts Northwestern Switzerland, 5210, Windisch, Switzerland}

\begin{abstract}
Much evidence suggests that the solar corona is heated impulsively, meaning that nanoflares may be ubiquitous in quiet and active regions (ARs). Hard X-ray (HXR) observations with unprecedented sensitivity $>$3~keV are now enabled by focusing instruments. We analyzed data from the \textit{Focusing Optics X-ray Solar Imager (FOXSI)} rocket and the \textit{Nuclear Spectroscopic Telescope Array (NuSTAR)} spacecraft to constrain properties of AR nanoflares simulated by the EBTEL field-line-averaged hydrodynamics code. We generated model X-ray spectra by computing differential emission measures for homogeneous nanoflare sequences with heating amplitudes $H_0$, durations $\tau$, delay times between events $t_N$, and filling factors $f$. The single quiescent AR observed by \textit{FOXSI-2} on 2014 December 11 is well fit by nanoflare sequences with heating amplitudes 0.02 erg cm$^{-3}$ s$^{-1}$ $<$ $H_0$ $<$ 13 erg cm$^{-3}$ s$^{-1}$ and a wide range of delay times and durations. We exclude delays between events shorter than $\sim$900 s at the 90\% confidence level for this region. Three of five regions observed by {\nustar} on 2014 November 1 are well fit by homogeneous nanoflare models, while two regions with higher fluxes are not. Generally, the {\nustar} count spectra are well fit by nanoflare sequences with smaller heating amplitudes, shorter delays, and shorter durations than the allowed \textit{FOXSI-2} models. These apparent discrepancies are likely due to differences in spectral coverage between the two instruments and intrinsic differences among the regions.  Steady heating ($t_N$ = $\tau$) was ruled out with $>$99\% confidence for all regions observed by either instrument. 
\end{abstract}


\section{Introduction}
It has been known for nearly eighty years that the solar corona is significantly hotter than the solar photosphere \citep{Gro1939, Edl1943}. However, a complete explanation of this temperature gap has been difficult to achieve. While significant progress has been made in recent years, it is still unclear what the energetic contributions of different physical mechanisms such as waves, reconnection, and spicules are \citep{Kli2015,Par2012}. 

Two primary physical mechanisms are thought to contribute to high coronal temperatures: magnetic reconnection of stressed field lines and dissipation of MHD waves. Both involve heating on timescales much smaller than the cooling time of individual magnetic strands, and can therefore be characterized as impulsive heating \citep{Kli2006}. \citet{Par1988} coined the term ``nanoflare'' to describe magnetic reconnection between individual flux tubes, a process that can lead to subsequent heating and particle acceleration. However, the term is now widely used to describe impulsive heating events acting on individual flux tubes, in which cooling timescales are longer than heating timescales, without any preference for physical mechanism.  As pointed out by \citep{Kli2006}, all plausible mechanisms of coronal heating under realistic conditions predict that the heating is impulsive. This includes wave heating, whether the waves are dissipated by resonance absorption, phase mixing, or Alfvenic turbulence.

Nanoflares can be characterized by their volumetric heating amplitude $H_0$, duration $\tau$, and characteristic delay time between events $t_N$. A significant amount of research has focused on the nanoflare heating frequency (1/$t_N$) and how it compares to the characteristic cooling time $t_{cool}$ of a loop strand. High-frequency heating occurs for $t_N << t_{cool}$, while low-frequency heating occurs for $t_N >> t_{cool}$. Steady heating is simply the limit as $t_N$ approaches 0. If low-frequency nanoflares are prevalent, they will produce hot ($\ge$5~MK) plasma throughout the solar corona. However, emission at these temperatures is difficult to detect directly for two reasons: only small amounts of this plasma are predicted, and ionization non-equilibrium can prevent the formation of spectral lines that would form at those temperatures under equilibrium conditions \citep{Gol1989, Bra2006, Rea2008, Bra2011}.   

Field-aligned and field-line-averaged hydrodynamic simulations have been used to predict the differential emission measure distributions DEM(T) = $n^2 dh/dT$ produced by nanoflares with a wide range of physical properties \citep{Car2014, Bar2016a, Bar2016b}. Here $n$ is the plasma density, and $dh/dT$ corresponds to spatial variations in the temperature field along a particular line of sight. In addition, the DEM distributions of active regions have been measured by extreme ultraviolet (EUV) and soft X-ray (SXR) instruments including the \textit{Solar Dynamics Observatory's} Atmospheric Imaging Assembly (AIA, \citealt{Lem2012}), the \textit{Hinode} X-Ray Telescope (XRT, \citealt{Gol2007}) and the \textit{Hinode} EUV Imaging Spectrometer (EIS, \citealt{Cul2007}). In general these distributions peak close to 4 MK and fall off steeply at higher and lower temperatures \citep{Tri2011, War2012, Sch2012}. \citet{Car2014} and \citet{Car2015} found, through large numbers of simulations, that nanoflare sequences with delay times of hundreds to $\sim$2000 s ($t_N \sim t_{cool}$) give results that are consistent with AR observations. In addition, these studies found that delay times proportional to the total nanoflare energy are required to match the broad range of $EM$ slopes found in previous studies. \citet{Bra2016} created model active regions heated by nanoflares and showed that the best agreement with AR observations occurs for delay times on the order of a loop cooling time (several thousand seconds). Time-lag measurements of ARs at multiple wavelengths have shown signs of widespread cooling and are also consistent with $t_N$ values on the order of several thousand seconds \citep{Via2012, Via2017}. While active region observations with AIA, XRT, and EIS can strongly constrain AR emission below $\sim$5~MK, constraints are less stringent at higher temperatures \citep{Win2012}.   

Hard X-ray (HXR) instruments can be used to detect or constrain plasma at temperatures $\gtrsim$5~MK. HXR emission is not sensitive to ionization non-equilibrium effects, which can suppress line emission from high-temperature plasmas. However, such plasma can still be difficult to detect because the temperature of a cooling, post-nanoflare flux tube peaks well before the luminosity (which is proportional to the DEM in a given temperature bin). Searches for hot plasma from nanoflares have been performed during periods of low solar activity, in order to avoid contamination from resolvable flares. Long duration, spatially-integrated observations from the \textit{Reuven Ramaty High Energy Solar Spectroscopic Imager} ({\rhessi}, \citealt{Lin2002}) the \textit{Solar PHotometer IN X-rays} (\textit{SphinX}, \citealt{Syl2008}), the X-123 spectrometer and the EUNIS rocket experiment have all shown evidence of plasma at $T>$5~MK during non-flaring times \citep{McT2009,Mic2012,Cas2015, Bro2014}. The combination of XRT and {\rhessi} was used to set constraints on a high-temperature component in active regions by \citet{Rea2009} and \citet{Sch2009}. Large uncertainties in these analyses prevented a definitive detection; although {\rhessi} is more responsive to high-temperature plasma than the instruments on \textit{Hinode}, it lacks the sensitivity to reliably obtain images and spectra from non-flaring active regions. 

Improved sensitivity and dynamic range can be obtained at energies $>$3~keV by the use of HXR focusing optics. This technology has enabled direct imaging of HXR photons in place of the indirect images obtained by previous instruments such as {\rhessi}. The \textit{Focusing Optics X-ray Solar Imager (FOXSI)} sounding rocket payload uses focusing optics to image the Sun with much higher sensitivity and dynamic range than {\rhessi} \citep{Gle2016}. {\foxsi} has flown twice (in 2012 and 2014) and is expected to fly again in 2018. The \textit{Nuclear Spectroscopic Telescope Array (NuSTAR)} is a NASA Astrophysics Small Explorer launched on 2012 June 13 \citep{Har2013}. While it was not designed to observe the Sun, {\nustar} has successfully done so on thirteen occasions without any damage to the instrument; for a summary of the first four solar pointings see \citealt{Gre2016}. Both {\foxsi} and {\nustar} have been used to perform imaging spectroscopy of active regions and to set limits on hot plasma in those regions \citep{Ish2014, Han2016, Ish2017}.  

\begin{figure}[htp]
  \centering
  \includegraphics[width=\columnwidth]{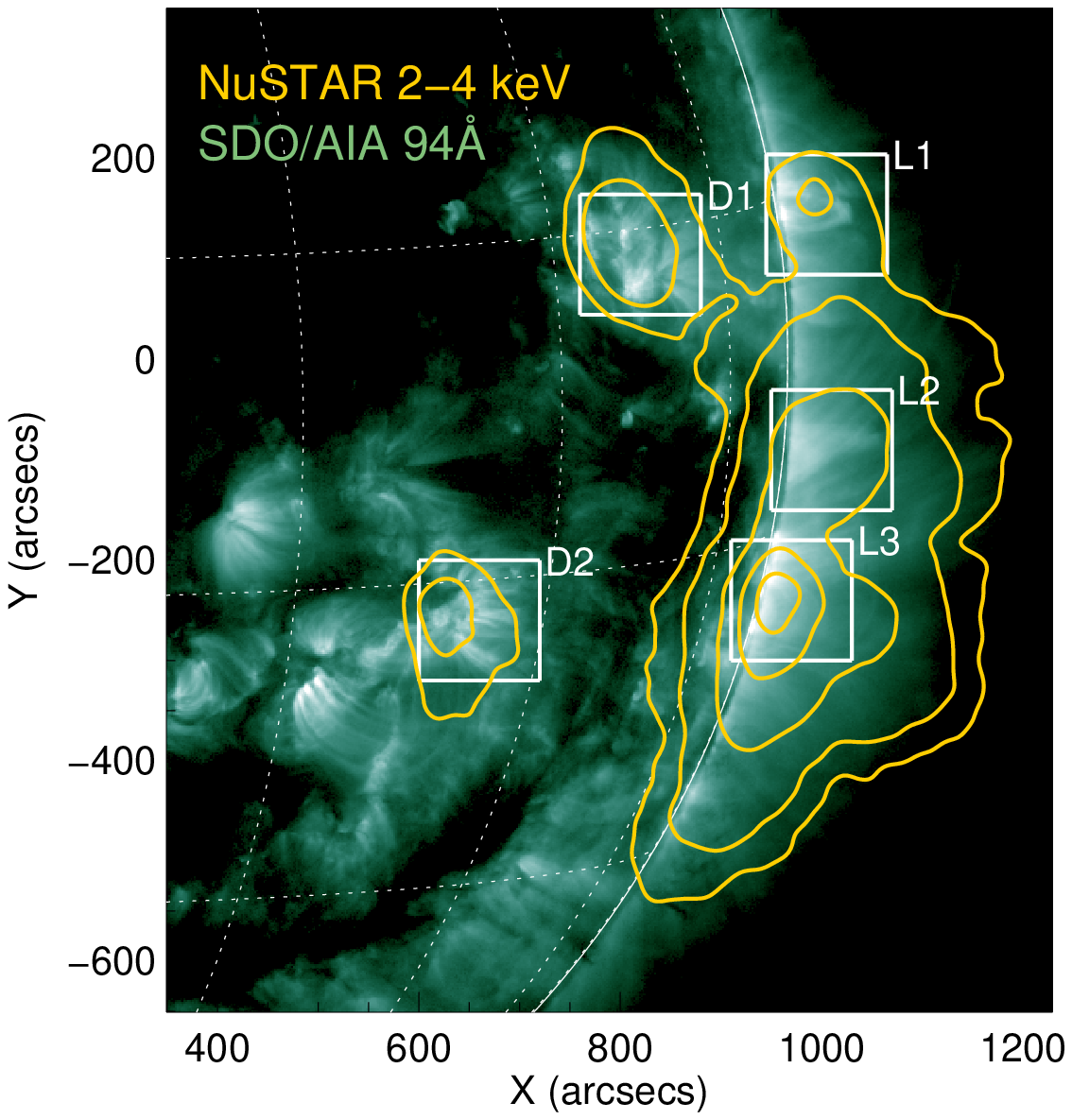}
  \includegraphics[width=\columnwidth]{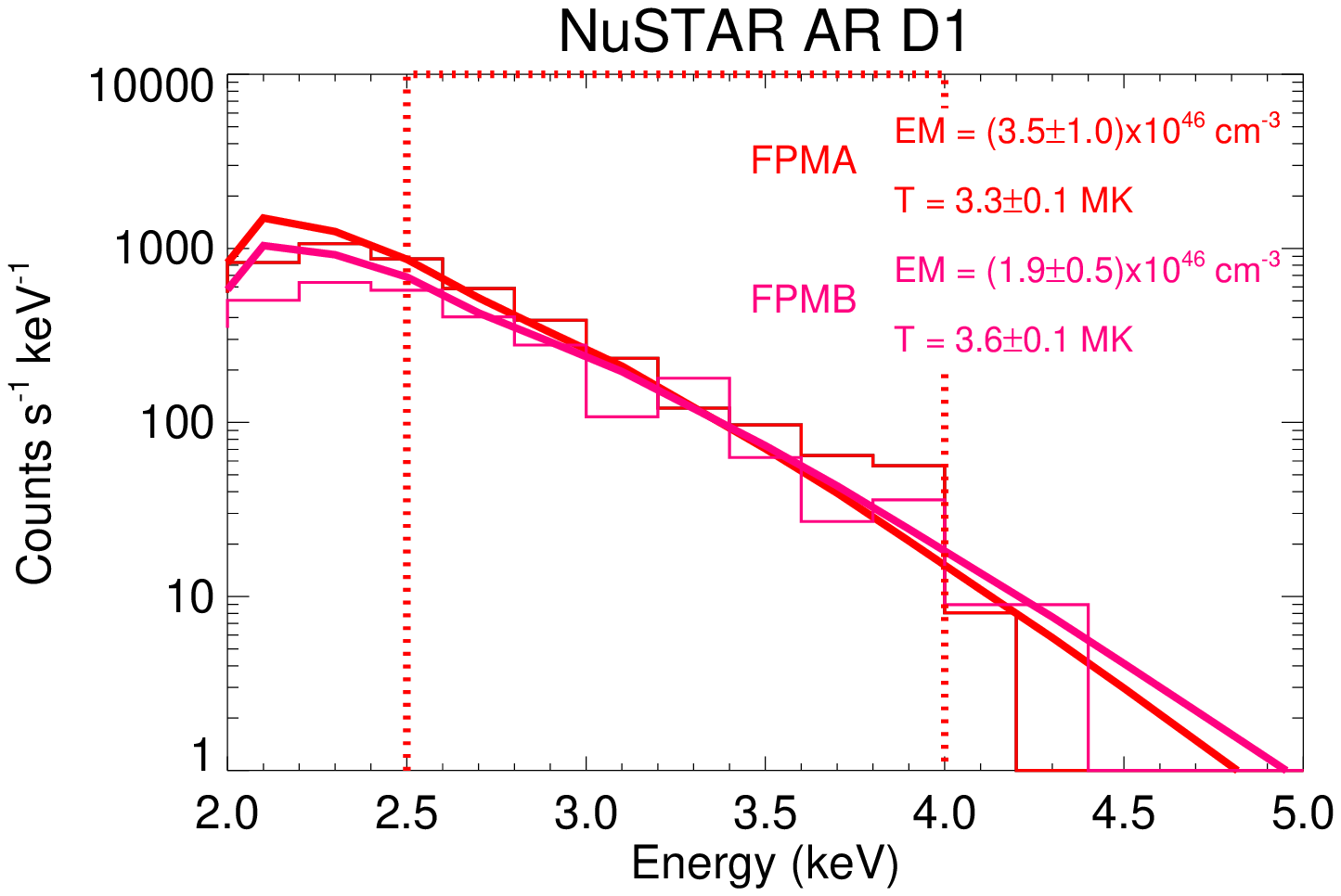}
\caption{(Top) Combined EUV and HXR image of five active regions observed by {\nustar} on 2014 November 1, with an effective HXR exposure time of 3.11 s. {\nustar} 2--4~keV flux contours (5, 10, 25, 50, and 80\%) from the FPMA telescope are overlaid in yellow on a co-temporal AIA 94~$\angstrom$ image. The {\nustar} image is co-aligned with the AIA data and smoothed (7$''$ Gaussian smoothing). White boxes are the areas used for this analysis.  (Bottom) {\nustar} count spectra from the FPMA and FPMB telescopes for one of the on-disk active regions (D1) observed on 2014 November 1. The fit energy range is shown by the dashed box. Isothermal fit parameters and uncertainties are given in the upper right corner.  As shown in this paper, there are a wide variety of energy distributions (going far beyond this isothermal model) that can well fit these data.}
  \label{fig:nustar_image_spec}
\end{figure} 

In this paper we use active region observations from {\nustar} and \textit{FOXSI-2} to constrain the physical properties of nanoflares, particularly their heating amplitudes, durations, and delay times. We utilize {\nustar} and \textit{FOXSI-2} datasets that were analyzed in \citet{Han2016} and \citet{Ish2017}, respectively. We describe solar observations with these instruments in $\S$\ref{observing}, discuss our analysis methods in $\S$\ref{methods}, present our results in $\S$\ref{discussion}, and describe our conclusions and future work in $\S$\ref{conclusions}. 

\section{Solar Observations with {\nustar} and {\foxsi}}
\label{observing}
{\nustar} has two co-aligned X-ray optics focused onto two focal plane detector arrays (FPMA \& FPMB), with a field-of-view of $\sim$12$'$$\times$12$'$ and a half-power diameter of $\sim$65$''$ \citep{Mad2015}. {\nustar} is well calibrated over the 3--79~keV bandpass, and the lower energy bound can be extended to 2.5 keV if there is sufficient flux present. {\nustar} has successfully observed active regions \citep{Gre2016,Han2016,Kuh2017}, the quiet Sun \citep{Mar2017}, and small (GOES class $<$A1) solar flares \citep{Gle2017, Wri2017, kuhar2018} with unprecedented sensitivity. Summary plots of all {\nustar} observations can be found at \url{https://ianan.github.io/nsigh_all/}. Of particular interest to us are quiescent active region observations on 2014 November 1, described in detail by \citet{Han2016}. Figure \ref{fig:nustar_image_spec} shows {\nustar} 2--4~keV contours overlaid on a co-temporal AIA 94 $\angstrom$ image of five active regions seen during this campaign. Two of the observed regions (D1 and D2) were fully on-disk, while the other three (L1, L2, and L3) were partially occulted. Count spectra from both {\nustar} telescopes, as well as the corresponding isothermal fits, are shown in Figure \ref{fig:nustar_image_spec} for one of these regions (D1). The other ARs had isothermal fit temperatures from 3--4.5~MK and emission measures from 10$^{46}$--10$^{47}$ cm$^{-3}$. 

\begin{figure}[htbp]
\centering
\includegraphics[width=\columnwidth, trim=0 0.5cm 1.5cm 1.5cm, clip=true]{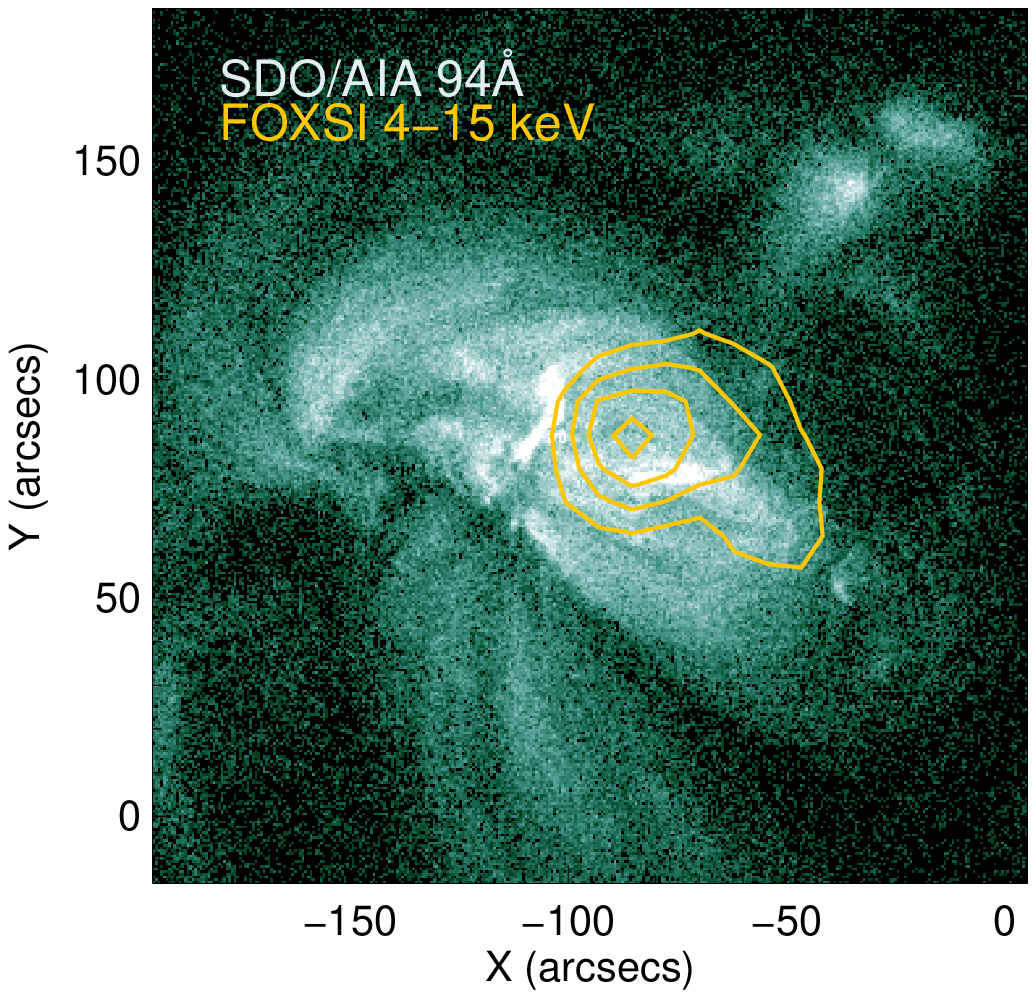}
  \includegraphics[width=\columnwidth]{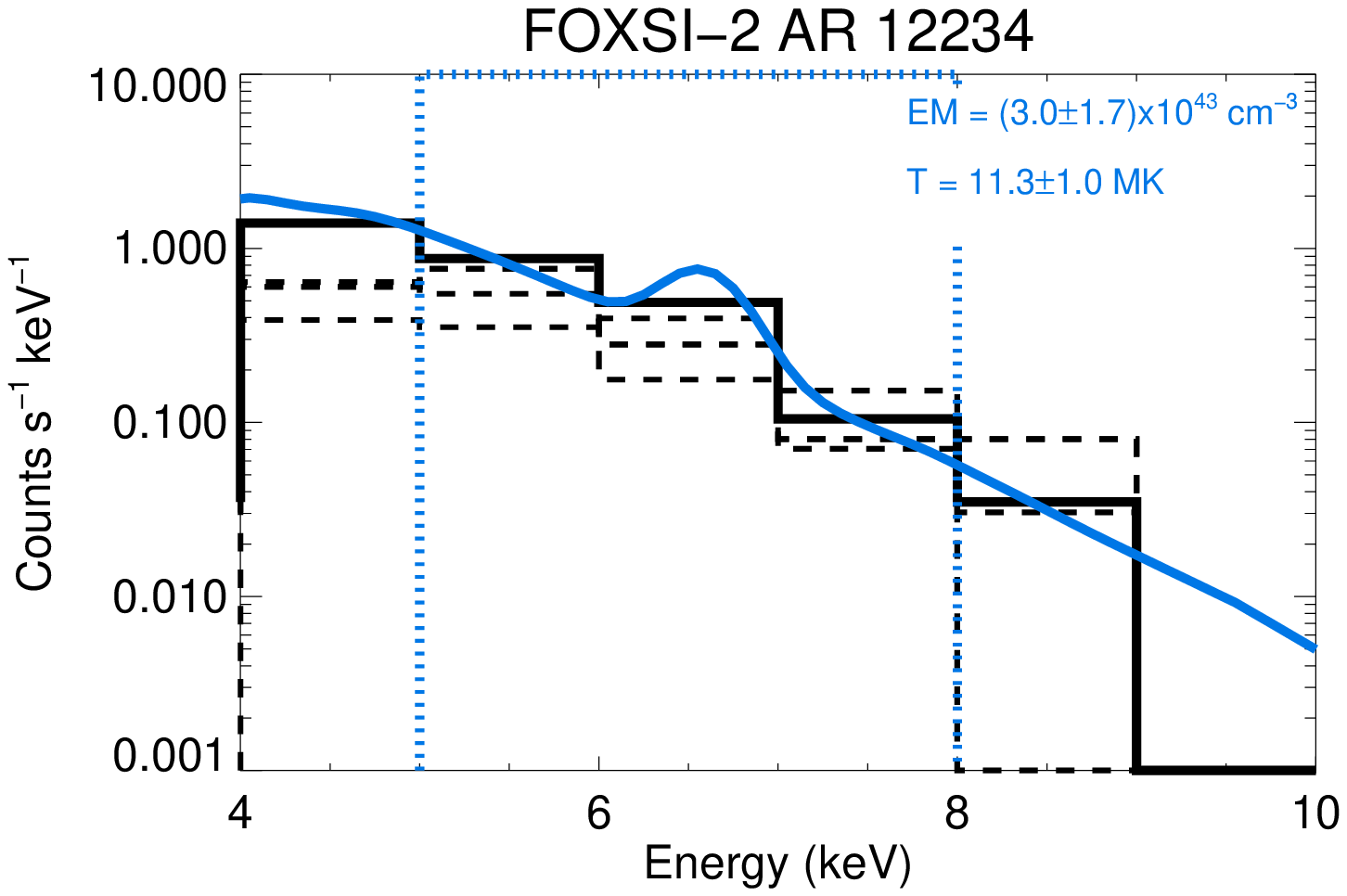}
\caption{(Top) \textit{FOXSI-2} 4--15~keV  HXR contours from Det 6 overlaid on a co-temporal AIA image of AR 12234. The \textit{FOXSI-2} contours have been chosen to show 30, 50, 70, and 90\% of the maximum value, and the \textit{FOXSI-2} effective exposure time is 38.5 s.  (Bottom) \textit{FOXSI-2} count spectra of AR 12234 from 4 Si detectors; the Det 6 spectrum is plotted as a solid line and the Det 0, Det1, and Det 5 spectra are plotted with dashed lines. (The optic/detector pairs have different responses.) The best-fit isothermal T, EM, and 1-sigma uncertainties for the Det 6 spectrum are written on the plot, and the fit range is marked by the dashed box. This spectrum was integrated over an exposure time of 38.5 s.  As shown in \citet{Ish2017}, a multithermal model gives a better fit than this isothermal approach when considering \textit{FOXSI} and \textit{Hinode}/XRT data combined.}
\label{fig:foxsi_image_spec}
\end{figure}

\textit{FOXSI} is a sounding rocket payload that uses focusing optics to directly image solar photons between 4--20~keV. \textit{FOXSI} has flown twice from White Sands, New Mexico and has observed small solar flares, active regions, and the quiet Sun. We analyzed non-flaring AR data from the second \textit{FOXSI} flight on 2014 December 11 \citep{Gle2016}. \textit{FOXSI-2} targeted several areas of the Sun during the course of its 6.5 minute flight, including an active region near disk center (NOAA AR 12234) that was quiescent for the duration of this observation. Figure \ref{fig:foxsi_image_spec} shows \textit{FOXSI-2} 4--15~keV contours integrated over the exposure time (38.5 s) and overlaid on a co-temporal AIA 94 $\angstrom$ image. Also shown is a \textit{FOXSI-2} count spectra of AR 12234 with 1.0~keV bins integrated over the observing period. Data from four Si detectors (Det 0, Det 1, Det 5, and Det 6) are included in this figure. The spectrum from the detector with the greatest response (Det 6) is fit well by an isothermal plasma with temperature T = 11.3~MK and emission measure EM = 6.0$\times$10$^{43}$ cm$^{-3}$, at a reduced chi-squared value of 0.95. While the count fluxes from this active region are fairly low, there is clear evidence for the presence of plasma $\gtrsim$10~MK within the uncertainties of the spectral fit. The iron line complex at 6.7~keV is a well-known indicator of temperatures above $8$~MK \citep{Phi2004}. A full differential emission measure (DEM) analysis of this active region with \textit{FOXSI-2} and \textit{Hinode} has been performed by \citet{Ish2017}. That paper uses multi-wavelength observations to provide the most direct detection to date of $>$10~MK plasma in a non-flaring solar active region. In this work, we attempt to characterize the impulsive heating parameters that may have produced this emission.

We wish to emphasize that we start with isothermal fits only to show the traditional way of analyzing HXR data, and to emphasize the different sensitivities of the two instruments.  In general, we do not expect these active regions to contain only a single temperature, as there is a broad base of literature finding multithermal distributions in active regions.  Furthermore, the \textit{FOXSI}-2 active region has been demonstrated by \citet{Ish2017} to be multithermal when considering \textit{Hinode}/XRT data alongside the \textit{FOXSI}-2 data; temperatures of at least 3--15 MK were found.  An isothermal fit to a multithermal temperature distribution picks out the temperature to which the instrument is the most sensitive.  The very different temperatures found by \textit{FOXSI}-2 and \textit{NuSTAR} for the two active regions could be due to intrinsic differences in the active regions themselves, or in the sensitivities of the two instruments, which measure peak rates in different energy ranges (2--2.5 keV for \textit{NuSTAR}; 4--5 keV for \textit{FOXSI-2}).  In this paper, we institute no constraint on the multithermal nature of the plasma and accept any nanoflare distribution that can well fit the observed data.

\begin{figure*}[htp]
  \centering
  \includegraphics[width=\textwidth]{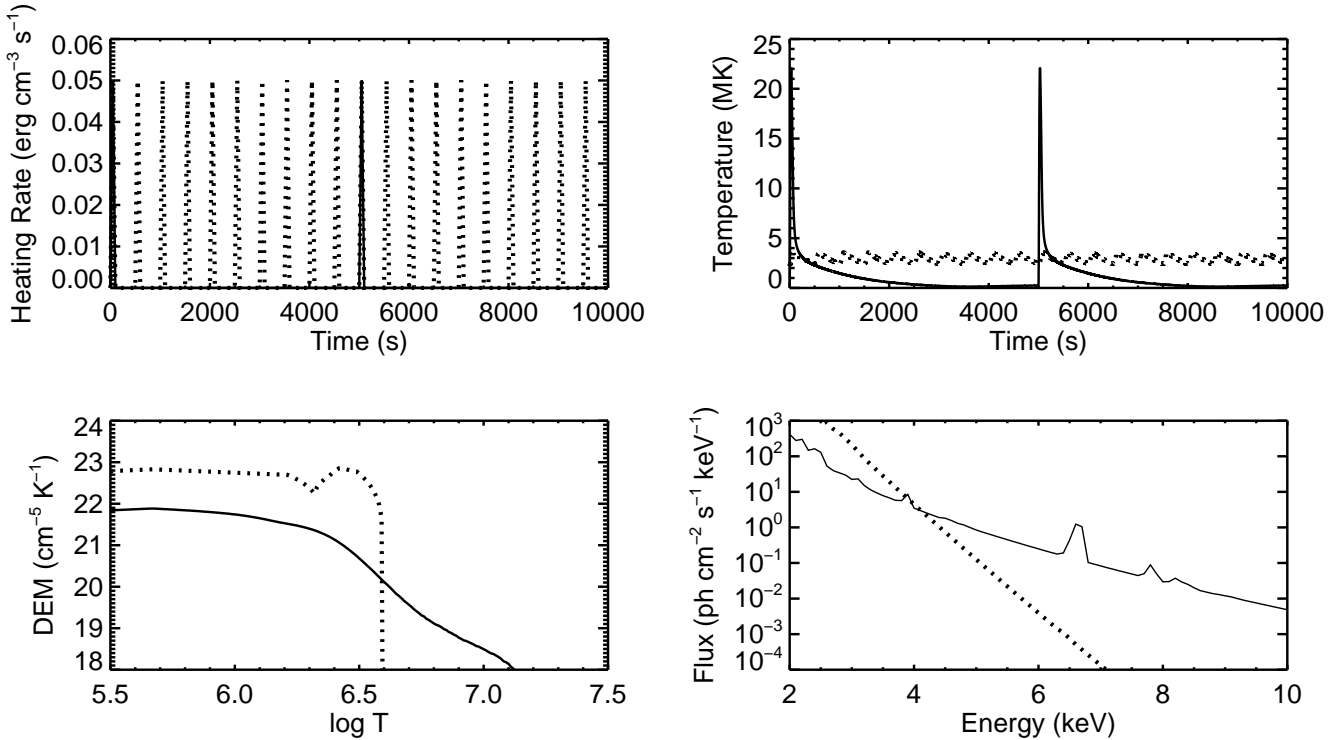}
  \caption{EBTEL simulations of high-frequency ($t_N = 500$ s) and low-frequency ($t_N = 5000$ s) nanoflare heating in a single loop strand with H$_0$ = 0.05 erg cm$^{-3}$ s$^{-1}$, $\tau$ = 100 s, and L = 2$\times$10$^{9}$ cm. Low-frequency values are indicated with solid lines and high-frequency values with dashed lines. Both nanoflare sequences were started 10000 s before the plotted times to erase the initial plasma conditions. (Top left) Volumetric heating rate as a function of time. (Top right) Average loop temperature as a function of time. (Bottom left) DEM distributions time-averaged over the last nanoflare cycle of each sequence. The discontinuity in the high-frequency curve is the intersection of the coronal and TR DEM distributions. (Bottom right) Simulated X-ray spectra derived from the time-averaged DEMs and integrated over a 60$\times$60 arcsecond$^{2}$ area.}
  \label{fig:nanoflares}
\end{figure*}

\section{Methods}
\label{methods}
\subsection{Physical Parameters and Their Selection}
We simulated homogeneous nanoflare sequences, in which every nanoflare is identical and evenly spaced, with the Enthalpy-Based Thermal Evolution of Loops (EBTEL) field-line-averaged hydrodynamics code \citep{Kli2008,Car2012a,Car2012b}. EBTEL is widely used in the solar physics community, and model outputs have been benchmarked against field-aligned numerical codes such as HYDRAD \citep{Bra2013}. An updated version, ebtel++\footnote{\url{https://rice-solar-physics.github.io/ebtelPlusPlus/}}, improves upon the original IDL code by incorporating two-fluid hydrodynamic equations and modifying certain parameters to produce better agreement with field-aligned simulations \citep{Bar2016a}. The new code also provides an adaptive timestep routine that ensures the timestep is always sufficiently small compared to the timescales of the relevant physical processes (for more details, see the appendices of \citealt{Bar2016a}). Subsequently, for short heating timescales and large heating rates ebtel++ is more accurate. It also runs faster than the IDL code, and significantly reduced our computing time. When we refer to ``EBTEL'' hereafter we are referring to ebtel++. In our simulations only the electrons are heated; future work will include ion heating, as in \citet{Bar2016a}.

EBTEL accepts a user-defined time array, heating function (a homogeneous nanoflare sequence for this analysis), and loop half-length L as inputs, then subsequently calculates the loop-averaged pressure, density, and temperature at each time step. The input heating is the field-line-averaged volumetric heating rate. We note that the spatial dependence of the heating is not generally important, since coronal thermal conduction and flows are so efficient at spreading the energy along field lines.  EBTEL also computes the differential emission measure separately in the transition region (TR) and corona, for a loop strand with cross-sectional area A = 1 cm$^{2}$. This area is a default area for the computation and is not the actual area of a loop or strand.  We chose to use a triangular heating function for all our simulations. The pulse height is the heating amplitude $H_0$ in erg cm$^{-3}$ s$^{-1}$ and the width is the event duration $\tau$ in seconds. The delay $t_N$ is the time between the start of each heating event. In addition, we included a constant, low-level background heating of 3.5$\times$10$^{-5}$ erg cm$^{-3}$ s$^{-1}$ in every simulation. This term prevents catastrophic cooling of the loop strand at late times \citep{Car2013}, and is small enough that it otherwise has no effects on our results.   The background heating on its own heats the region to only $<$300,000 K and cannot account for the few or several million degree temperature of the active region.

Figure \ref{fig:nanoflares} shows heating functions and the corresponding temperature evolution, time-averaged DEMs, and HXR spectra for nanoflare sequences with $t_N$ = 500 s (high-frequency) and $t_N$ = 5000 s (low-frequency) occuring on a loop strand with a half-length L = 2$\times$10$^{9}$ cm. Low-frequency heating results in a DEM that extends to higher temperatures and a harder photon spectrum compared to high-frequency heating. This is because low-frequency heating gives the loop strand more time to cool and drain before the next event. The lower density at the time of the next event means that the plasma can be heated to a higher temperature. Note that, not only do high-frequency nanoflares produce lower average temperatures for the same average heating rate, but even for events with the same heating amplitude and duration as shown in Figure 5. Here the high-frequency nanoflare sequence contains an order of magnitude higher average heating rate than the low-frequency case.  

\begin{table}
  \centering
   \begin{tabular}{ | p{4cm} | p{4cm} | }
   \hline
   Active Region & Loop Half-Length (cm) \\ \hline
    AR 12234 & 6$\times$10$^{9}$ \\ \hline
    NuSTAR D1 & 7$\times$10$^{9}$ \\ \hline
    NuSTAR D2 & 7$\times$10$^{9}$ \\ \hline
    NuSTAR L1 & 7$\times$10$^{9}$ \\ \hline
    NuSTAR L2 & 1$\times$10$^{10}$ \\ \hline
    NuSTAR L3 & 7$\times$10$^{9}$ \\
    \hline
   \end{tabular}
    \caption{Table of estimated loop lengths for the five {\nustar} and single \textit{FOXSI-2} active regions. These lengths were calculated from the manual selection of loop footpoints in AIA 171~$\angstrom$ images.}
    \label{tab:loop_lengths}
\end{table}

The physical parameters that alter the X-ray spectrum are $H_0$, $\tau$, $t_N$, L, and the filling factor $f$, a normalization that reflects the fact that in a given volume of the corona, only a certain fraction of loop strands may be impulsively heated. We varied $H_0$, $\tau$, and $t_N$ across a range of values for each active region to determine which parameter combinations gave good agreement with observations. For each set of parameters we simulated a sequence of five nanoflares and used the DEM values from the last nanoflare cycle (starting with the heating event and ending after one delay time). We used only the last cycle in order to eliminate the initial EBTEL plasma conditions. The shortest value of delay was set to the longest value of duration to avoid overlapping events; quasi-continuous heating occurs when the delay and duration are exactly equal. In future work we will explore the effect of using non-homogeneous nanoflare sequences where, for example, the delay varies as a function of nanoflare energy. The average loop half-length L was estimated separately for each region with AIA images using the following procedure.
 
\begin{table}
  \centering
   \begin{tabular}{ | p{4cm} | p{4cm} | }
   \hline
   Physical Parameter & Range of Tested Values \\ \hline
    $H_0$ & 0.005--25 erg cm$^{-3}$ s$^{-1}$ \\ \hline
    $\tau$ & 5--500 s \\ \hline
    $t_N$ & 500--10,000 s \\ 
    \hline
   \end{tabular}
    \caption{Range of physical parameters for simulated nanoflare sequences.}
    \label{tab:params}
\end{table}

The \textit{FOXSI-2} observation of AR 12234 took place when this region was close to disk center. To estimate the average coronal loop length, we measured the distances between several visible pairs of loop footpoints in the AIA 171~$\angstrom$ channel. The regions observed by {\nustar} on 2014 Nov 1 were near or over the solar limb, which made it difficult to measure the entire loops. Therefore, we used AIA 171~$\angstrom$ images from 2014 October 28 to calculate footpoint distances for these regions. After we measured the average footpoint separations we corrected for projection effects by dividing each distance by cos($\lambda$), where $\lambda$ is the central longitude of each region. We assumed semi-circular loop geometries and determined the average half-lengths L=$\pi d / 4$, where d is the longitude-corrected average footpoint separation for a given region. The loop length estimates for each region are listed in Table \ref{tab:loop_lengths}.  

\begin{figure}
\centering
\includegraphics[width=0.7\columnwidth]{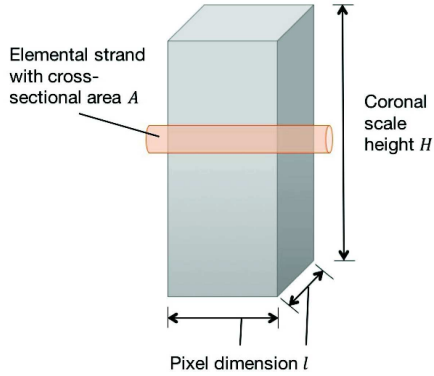}
\caption{This figure shows the geometry used to calculate the number of loop strands within a particular observing area, and subsequently to scale the simulated EBTEL DEM from a single strand. The horizontal strand approximation was made for the coronal portion only, and the transition region footpoints were treated separately (as shown in Equation \ref{eqn:dem}). }
\label{fig:dem_schematic}
\end{figure}

When looking at an active region through the optically thin corona, all the loops in various stages of heating and cooling along a line-of-sight contribute to each spatial pixel. Therefore we time-averaged the DEM distributions for the last cycle of each EBTEL simulation; this produced a superposition of every stage of heating and cooling in that cycle, similar to what we expect from observations. We assumed a fixed coronal scale height H = 5$\times$10$^{9}$ cm in order to calculate the number of loop strands in a volume with cross-sectional area equal to the area of a given action region. We then computed model photon spectra by first scaling each EBTEL (time-averaged) DEM to an expected DEM observation as follows: 
\begin{equation}
  \textrm{$\mathit{DEM}$}_{obs} = \frac{\ell^{2} H}{2L} <\textrm{$\mathit{DEM}$}_{cor}> + \frac{\ell^{2}}{2} <\textrm{$\mathit{DEM}$}_{tr}>
  \label{eqn:dem}
\end{equation}

Here DEM$_{cor}$ and DEM$_{tr}$ are the EBTEL time-averaged DEM distributions for the corona and transition region in cm$^{-5}$ K$^{-1}$, $\ell^{2}$ is the observing area in cm$^{2}$, H is the scale height, and L is the loop half-length for the AR of interest. The multiplicative factors for each term give the expected volumetric DEM$_{obs}$ (cm$^{-3}$ K$^{-1}$) in a rectangular region of length and width $\ell$, and the spatial approximation of horizontal strands going up to a height H is used (as shown in Figure \ref{fig:dem_schematic}) for the coronal portion of each strand. The DEM$_{tr}$ is divided by a factor of two so that the footpoint emission is not doubly counted, and is not scaled by H because the depth of the transition region is independent of the coronal scale height. 
  
The HXR spectrum was derived from DEM$_{obs}$ by determining the emission measure (EM, units of cm$^{-3}$) in each temperature bin of width log(T) = 0.01 between log(T) = 4.0 and log(T) = 8.5, and calculating the corresponding isothermal spectra. The resulting sum of every individual spectrum was then convolved with instrument response functions from either {\nustar} or \textit{FOXSI-2}. This allowed us to make straightforward comparisons to the observed count spectra for any set of model parameters. For on-disk regions such as AR 12234 and {\nustar} ARs D1 and D2, we expect a significant contribution from the transition region to the line-of-sight plasma emission and therefore used the sum of DEM$_{cor}$ and DEM$_{tr}$. For off-limb regions such as {\nustar} ARs L1, L2 and L3 we expect to see predominantly coronal emission. Therefore for L1, L2, and L3 we used DEM$_{cor}$ only. 

\begin{figure*}[htp]
\centering
    \begin{subfigure}
        \centering
        \includegraphics[width=0.7\linewidth]{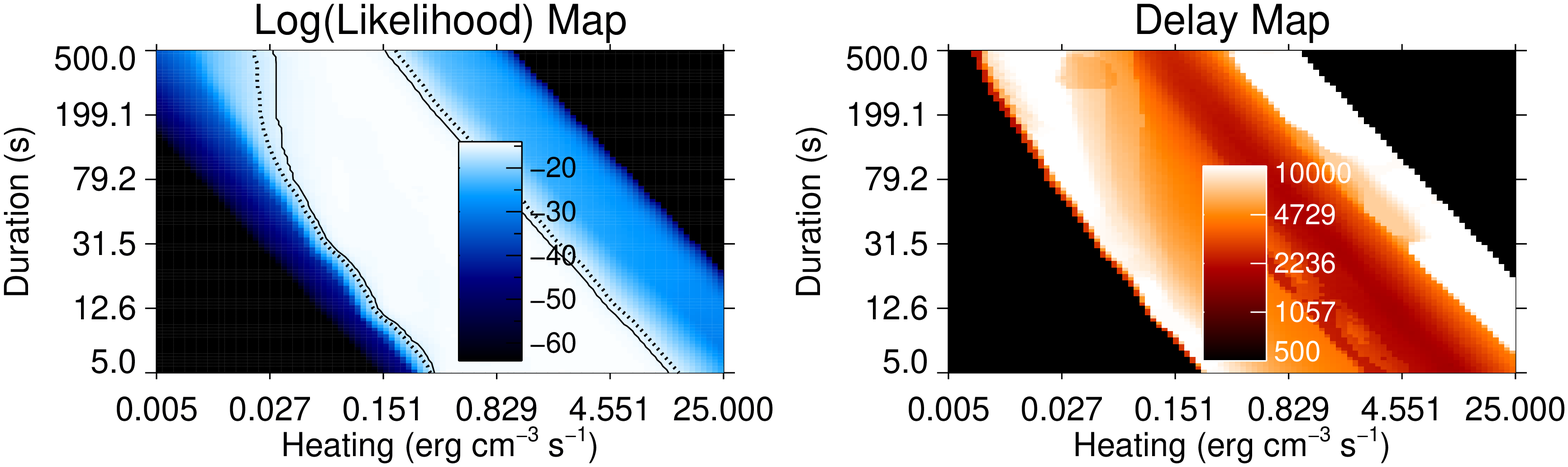} 
    \end{subfigure}
    \hfill
    \begin{subfigure}
        \centering
        \includegraphics[width=0.7\linewidth]{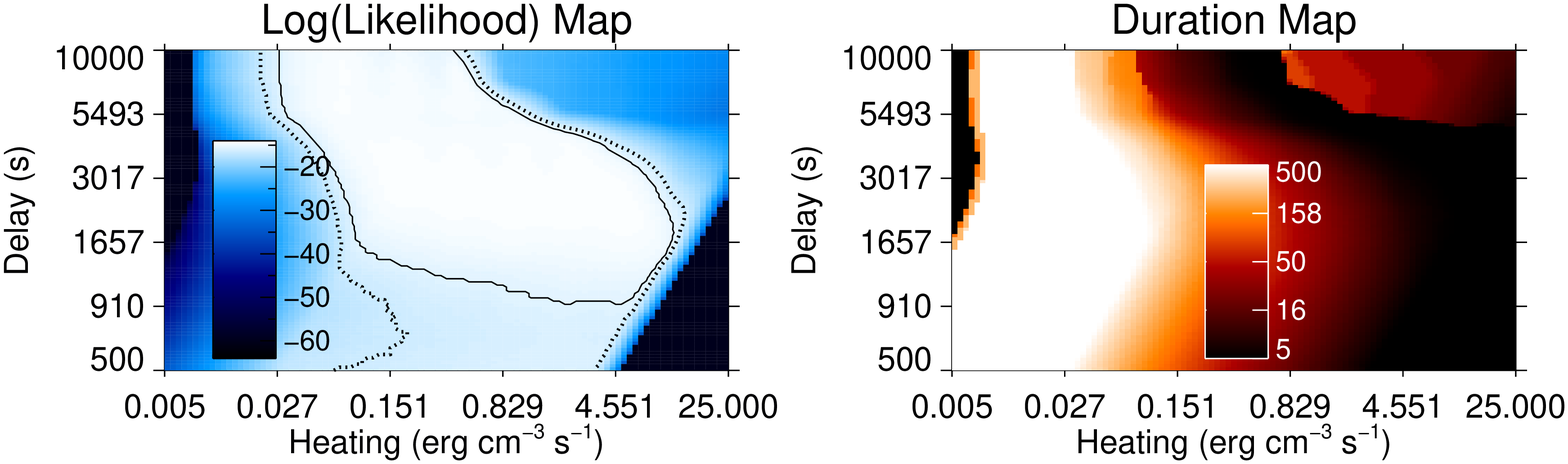}        
    \end{subfigure}
    \hfill
    \begin{subfigure}
        \centering
        \includegraphics[width=0.7\linewidth]{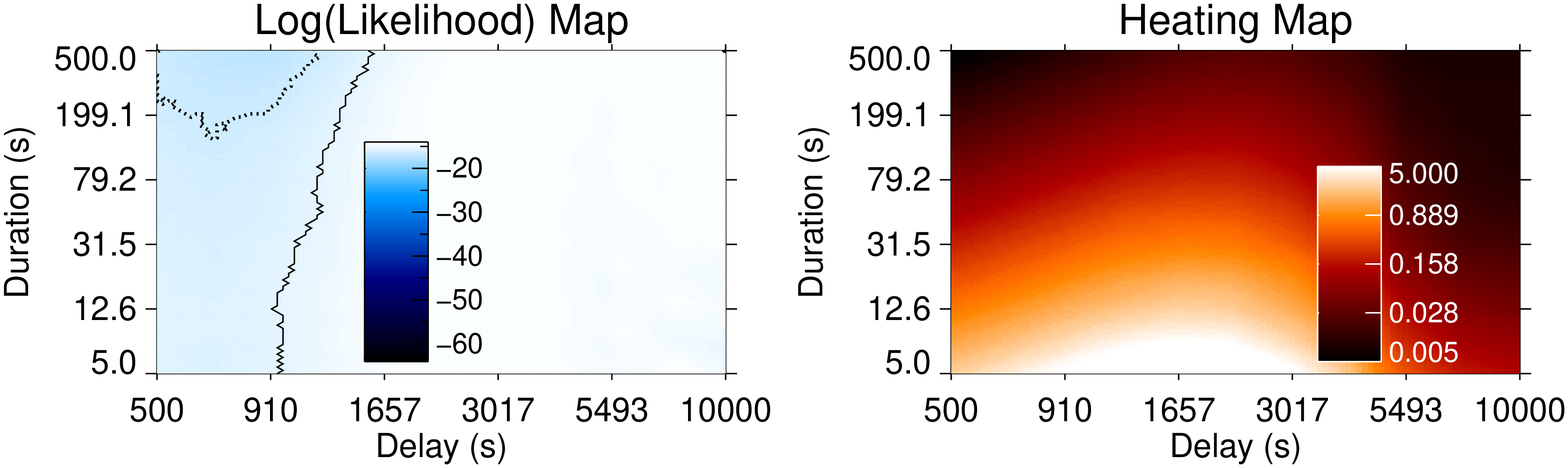}
    \end{subfigure}
  \caption{Parameter space results using combined data from four of the \textit{FOXSI-2} Si detectors (Det 0, Det 1, Det 5, and Det 6). (Left) 2D log likelihood intensity maps for each combination of $H_0$, $\tau$, and $t_N$. (Right) Intensity maps of the optimized third parameter corresponding to each 2D likelihood plot. Energy flux constraints (Equation \ref{eqn:flux_limit}) and EUV/SXR limits from AIA and XRT have been applied to the full parameter space. Both the likelihood and parameter maps were smoothed for display purposes using the procedure described in the text. Solid lines in the left panels show 90\% CIs and dotted lines show 99\% CIs for the case of 3 relevant parameters.} 
\label{fig:foxsi_all_limits}
\end{figure*}

We engaged in a systematic exploration of the nanoflare parameter space for each active region. Previous active region observations with EUV and SXR instruments are consistent with nanoflare delay times that range from hundreds to thousands of seconds \citep{Car2014} .  In the case of reconnection-related nanoflares, an event duration can be as short as the time that a reconnecting field line is in contact with a standing slow shock in the Petschek model, which is of order seconds \citep{Kli2006}. It could also be significantly longer (up to hundreds of seconds) if, for example, multiple reconnection events cluster together in space and time \citep{Kli2015}. The heating amplitude is not well-constrained theoretically, so we explored a wide range of values starting from a lower limit approximately two orders of magnitude above the background heating. The full range of physical parameters that we chose to explore is given in Table \ref{tab:params}. For every active region and instrument response, we created a 4D datacube with logarithmically spaced values of the nanoflare parameters $H_0$, $\tau$, and $t_N$ corresponding to the first 3 dimensions. The 4th dimension contained the model X-ray spectra from the EBTEL simulations corresponding to each set of parameter values. In order to reduce computational overhead we generated count spectra for an 11$\times$11$\times$11 array of $H_0$, $\tau$, and $t_N$, and then performed a 3D interpolation to obtain count spectra over an 101$\times$101$\times$101 array with the same minimum and maximum parameter values.  

We subsequently used the following procedure to generate 3D arrays containing the total likelihood for each active region and instrument response. The total likelihood is simply the product of individual likelihoods for a particular pair of modeled and observed count spectra \citep{Bev2003}. For these spectra the individual likelihoods are given by Poisson probabilities: 

\begin{equation}
  \mathcal{L} = \prod_{i=1}^{n} \mathcal{L}_i = \prod_{i=1}^{n} \frac{e^{-\mu_i} \mu_i^{x_i} }{x_i!}
\end{equation}

Here $\mu_i$ is the number of counts in the $i$th energy bin predicted by a particular nanoflare model and $x_i$ is the actual number of counts detected in that energy bin. Because both {\nustar} and \textit{FOXSI-2} count individual photons, we are free to choose our energy bins. The energy ranges we chose for these likelihood calculations were 2.5--5~keV for {\nustar} and 5--10~keV for \textit{FOXSI-2}, with bin widths of 0.2 and 1.0 keV respectively. We chose to use the likelihood statistic instead of chi-square because of the low number of counts in these ranges, including zero counts in some energy bins. For each combination of $H_0$, $\tau$, and $t_N$ we determined the value of the filling factor $f$ that resulted in the same cumulative number of counts in the modeled and observed spectra in the energy range of interest. This normalization of $f$ made it easier to determine what regions of parameter space for the physical quantities of primary interest ($H_0$, $\tau$, and $t_N$), resulted in the best agreement with observations. We calculated $\mu_i$ separately for response functions from the following instruments: the two {\nustar} telescopes (FPMA \& FPMB) and four \textit{FOXSI-2} Si detectors (Det 0, Det 1, Det 5, and Det 6). Then we computed total likelihood arrays for \textit{FOXSI-2} and {\nustar} by multiplying the individual detector arrays together. To visualize the parameter space we plotted 2D log likelihood intensity maps for every combination of $H_0$, $\tau$, and $t_N$. For every 2D coordinate pair (e.g. heating and duration), we determined the maximum likelihood in the 3rd dimension and the corresponding third parameter value (e.g. delay). 

In order to obtain parameter ranges that led to good agreement with the observed HXR data, we generated confidence intervals (CIs) for every 2D coordinate pair at 90\% and 99\% confidence levels \citep{Ney1937}. For a given confidence level $\alpha$, the CI represents values for the population parameter(s) such that if an infinite number of CIs were constructed, a fraction $\alpha$ would contain the true parameter value(s). In other words, there is an a priori probability $\alpha$ that a single CI will contain the true value of the parameter(s) of interest. Therefore a higher confidence level, e.g. 99\% versus 90\%, will lead to wider confidence intervals.  

In our explorations of this parameter space we found many sets of solutions that gave acceptable fits to the HXR data. This is not surprising given the multidimensional nature of the parameter space and the degeneracy between the various parameters (for example, increasing either the heating amplitude or the event duration increases the energy in a particular nanoflare and also increases the predicted X-ray flux). However, this degeneracy made it critical to use as many external constraints as possible. 

\subsection{Constraints on the Nanoflare Parameter Space}
It is generally accepted that mechanical motions in and below the photosphere are the ultimate drivers of coronal heating \citep{Kli2006}. The Poynting flux associated with flows stressing the footpoints of magnetic fields is given by
\begin{equation}
	F = \frac{1}{4\pi} B_{V}^{2} V_{h} tan(\theta) \hspace{1cm} \textrm{erg cm$^{-2}$ s$^{-1}$}
\end{equation}
where $B_{V}$ is the vertical field, $V_{h}$ is the horizontal velocity and $\theta$ is the field tilt angle. Typical values observed in active regions are $\sim$100 G and 1 km s$^{-1}$. \citet{Wit1977} calculated an average coronal energy loss of 10$^{7}$ erg cm$^{-2}$ s$^{-1}$ in active regions, which implies an average tilt angle $\theta$ $\sim$ 20 degrees. For a given loop strand we do not expect the time-averaged energy flux to exceed 10$^{8}$ erg cm$^{-2}$ s$^{-1}$, as this would imply significantly larger photospheric velocities and/or tilt angles, which can be ruled out observationally. This flux can be re-written in terms of the physical parameters of a nanoflare sequence: 
\begin{equation}
\label{eqn:flux_limit}
        F = \frac{ H_0 \tau L }{2 t_N} \hspace{1cm} \textrm{erg cm$^{-2}$ s$^{-1}$}
\end{equation}
Recall that $H_{0}$ is the nanoflare peak heating amplitude, $\tau$ is the nanoflare duration, L is the loop half-length, and $t_{N}$ is the delay between events. We implemented the requirement throughout our analysis that the energy flux F $<$ 10$^{8}$ erg cm$^{-2}$ s$^{-1}$. 

We placed additional constraints on the nanoflare parameter space using co-temporal observations from AIA and XRT. AIA data are available for the {\nustar} and \textit{FOXSI-2} observations on 2014 November 1 and 2014 December 11 respectively, while XRT data is only available for the 2014 December 11 \textit{FOXSI-2} flight. We obtained active region fluxes in DN s$^{-1}$ pixel$^{-1}$ for multiple AIA wavelengths (94, 131, 171, 193, 211, 335~$\angstrom$) and multiple XRT filters (Be-thick, Al-thick, Ti-poly, Al-mesh, Al-poly/Ti-poly, C-poly/Ti-poly, C-poly, Be-thin, Be-med, Al-med, Al-poly). DN (datanumber) is the native flux unit of both instruments, and is proportional to the number of electrons generated by photons incident on the CCD cameras of each telescope. For each nanoflare model we calculated predicted fluxes for the appropriate instrument response functions in every waveband. We required the predicted AIA and XRT fluxes to be $<$3 times the spatially-averaged fluxes for the chosen AR, and if this requirement was not met for every wavelength we excluded that model from our results. We did not set a lower limit on the EUV/SXR fluxes because additional populations of nanoflares (at higher frequencies, for example) could be present at temperatures below the {\nustar} and \textit{FOXSI-2} sensitivity. 

\begin{figure}
\centering
  \begin{minipage}[c]{\columnwidth}
  \includegraphics[width=\columnwidth]{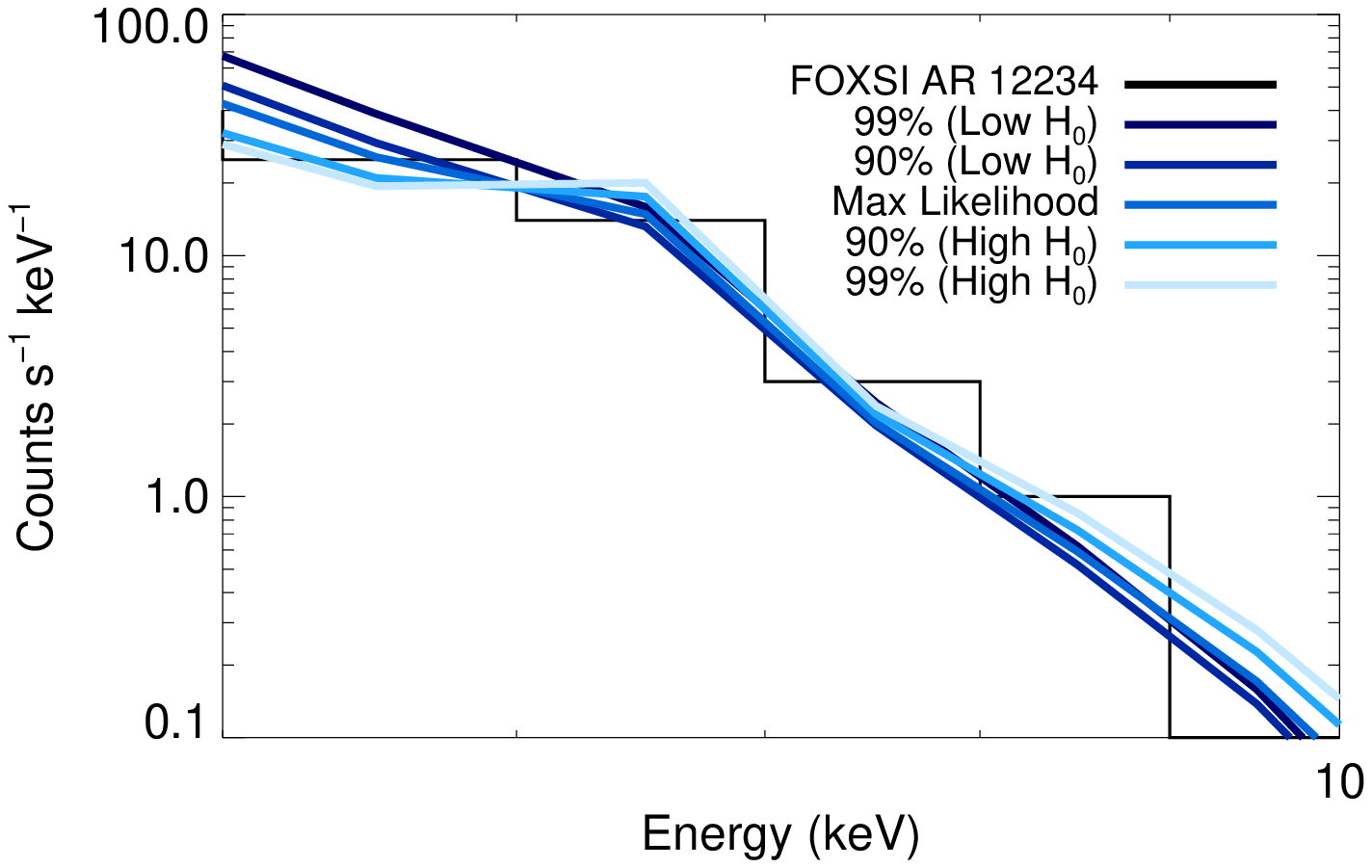}
  \end{minipage}\hfill
  \begin{minipage}[c]{\columnwidth}
	\footnotesize
	\begin{tabular}{|c|c|c|c|c|}
	\hline
					&	$H_0$	&	$\tau$ &	$t_N$	&	$f$	\\
					&	(erg~cm$^{-3}$~s$^{-1}$) & (s) & (s) & \\
	\hline
	99\% (Low $H_0$)	&	0.046	&	50	&	10000	&	0.62	\\
	90\% (Low $H_0$)	&	0.050	&	50	&	10000	&	0.42	\\
	Max Likelihood		&	0.46		&	50	&	3611		&	3.2$\times 10^{-4}$	\\
	90\% (High $H_0$)	&	1.27		&	50	&	2170		&	8.0$\times 10^{-6}$	\\
	99\% (High $H_0$)	&	1.50		&	50	&	2374		&	5.2$\times 10^{-6}$	\\
	\hline
	\end{tabular}
  \end{minipage}\hfill
\caption{\textit{FOXSI-2} Det 6 count spectrum of AR 12234 and predicted Det 6 spectra at five points in the optimized, constrained heating vs. duration parameter space (Figure \ref{fig:foxsi_all_limits}). For a fixed duration of $\tau$ = 50 s, we chose heating amplitudes at the maximum likelihood as well as on the 90\% and 99\% contours at lower and higher heating values.  The heating parameters corresponding to each curve are specified in the table.  }
\label{fig:foxsi_multimodel}
\end{figure}

\section{Results and Discussion}
\label{discussion}

\subsection{\textit{FOXSI-2} region}

Figure \ref{fig:foxsi_all_limits} shows 2D log likelihood and parameter intensity maps for \textit{FOXSI-2} observations of AR 12234, with the nanoflare models subjected to physical (energy flux) and observational (EUV/SXR) constraints. For each 2D coordinate pair (e.g. $H_0$, $\tau$), the third parameter (e.g. $t_N$) was chosen such that it maximized the likelihood. Before this optimization, a Gaussian smoothing kernel of width $\sigma$=1 pixel was applied to each 2D slice (101x101 pixels) of the 3D likelihood array in order to reduce visible interpolation artifacts. This also resulted in a slight smoothing of the parameter maps in the right panels. The black regions of parameter space in the two upper left panels ($H_0$ vs. $\tau$ and $H_0$ vs. $t_N$) are regions where the combination of energy flux and AIA/XRT constraints eliminated every value in the 3D array. The solid and dashed lines in the left panels indicate the 90\% and 99\% CIs, relative to the maximum likelihood, for three relevant parameters ($H_0$, $\tau$, $t_N$). \citet{Avn1976} showed that for three parameters of interest the 90\% (99\%) significance level is equivalent to an increase in the unreduced chi-square value of 6.25 (11.3) relative to the best fit. \citet{Wil1938} provided a mapping from chi-square to likelihood that allows us to plot likelihood significance levels: $-2 \textrm{log}(\mathcal{L}/\mathcal{L}_{max}) = \Delta \chi^{2}$. For 90\% CIs where $\Delta \chi^{2} = 6.25$, the likelihood level at which we draw contours is given by $\mathcal{L} = e^{-6.25/2} \mathcal{L}_{max} = 0.044 \mathcal{L}_{max}$; for 99\% CIs  $\mathcal{L} = e^{-11.3/2} \mathcal{L}_{max} = 0.0035 \mathcal{L}_{max}$.  

Figure \ref{fig:foxsi_multimodel} shows the \textit{FOXSI-2} AR 12234 count spectrum from Det 6 compared to five spectral models taken from the 2D heating/duration map. This figure shows the distinctions between models taken from points in parameter space at different confidence levels. We chose to sample nanoflare models at the maximum likelihood, as well as at lower and higher heating amplitudes on the 90\% and 99\% contours, for a fixed duration. 
The parameters for these sampled models are shown in the table below the spectrum.  

\begin{figure}[htp]
\centering
\includegraphics[width=\columnwidth]{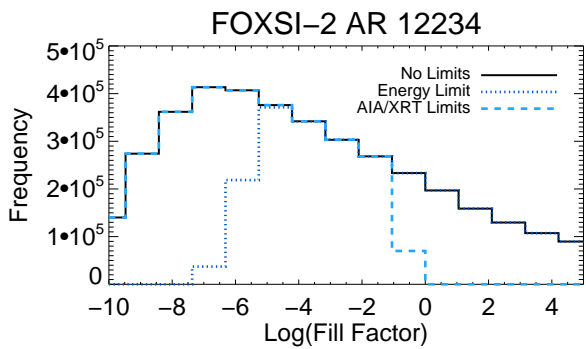}
\caption{Histograms of the fill factor for the \textit{FOXSI-2} AR and three different sets of constraints: no limits, energy flux limits, and AIA/XRT limits.}
\label{fig:foxsi_hist}
\end{figure}

\begin{figure*}[htp]
\centering
    \begin{subfigure}
        \centering
        \includegraphics[width=0.6\linewidth]{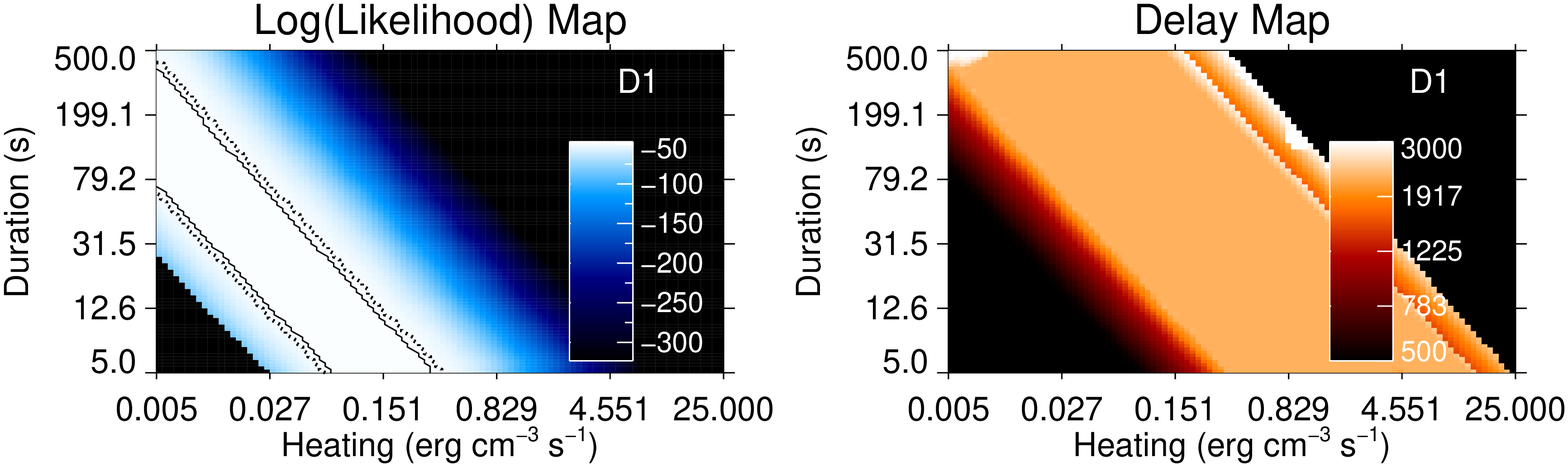} 
    \end{subfigure}
    \hfill
    \begin{subfigure}
        \centering
        \includegraphics[width=0.6\linewidth]{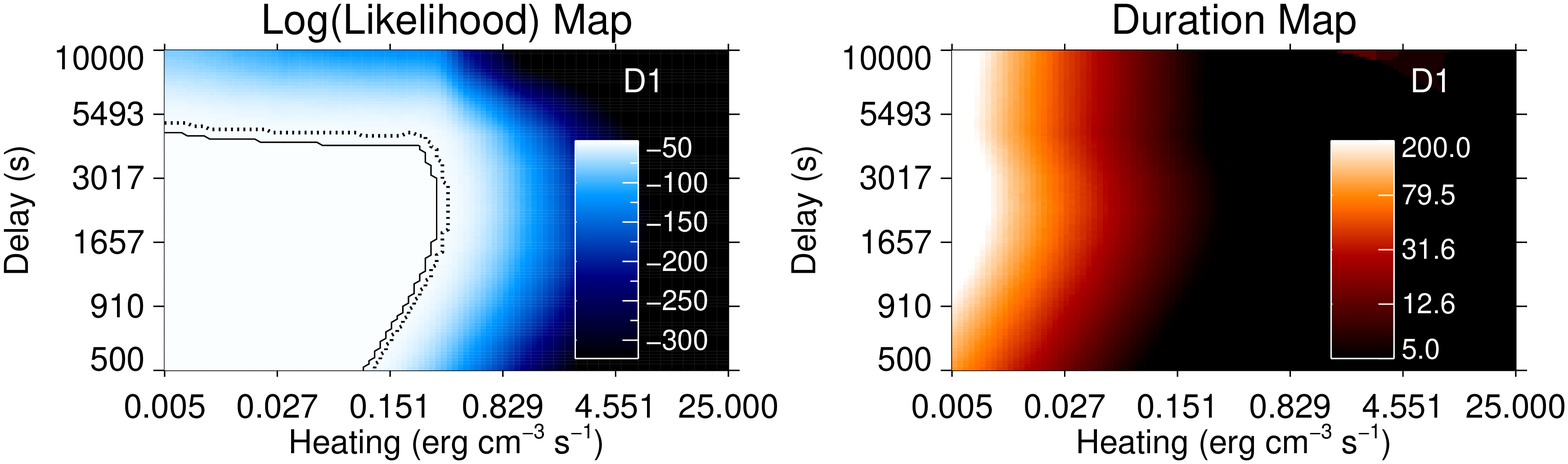}
    \end{subfigure}
    \begin{subfigure}
        \centering
        \includegraphics[width=0.6\linewidth]{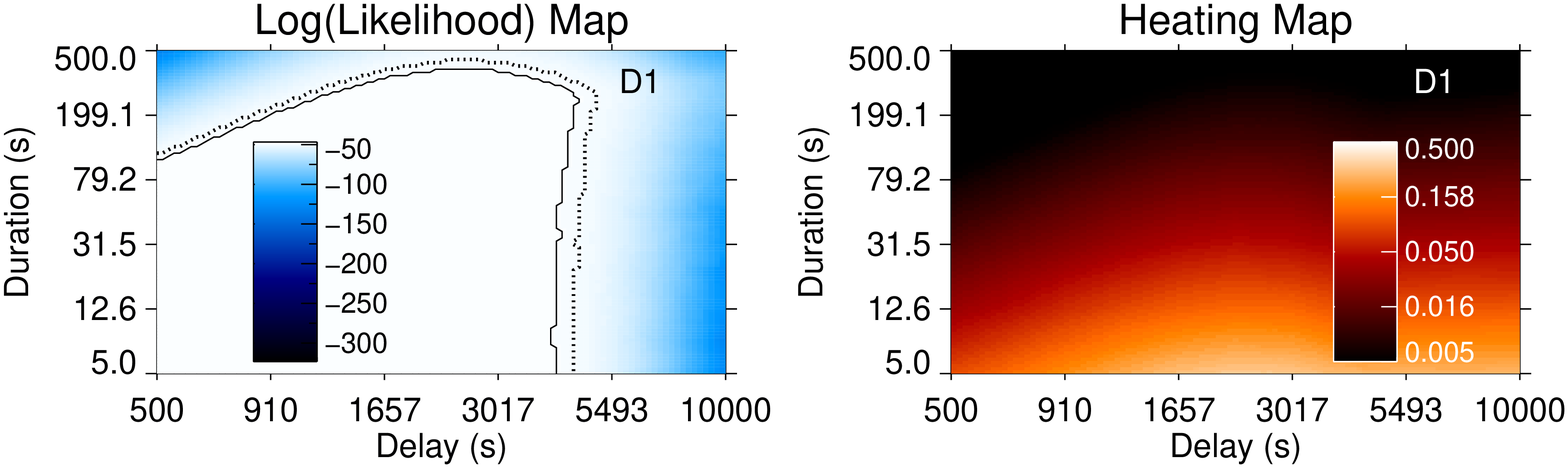} 
    \end{subfigure}
    \begin{subfigure}
        \centering
        \includegraphics[width=0.6\linewidth]{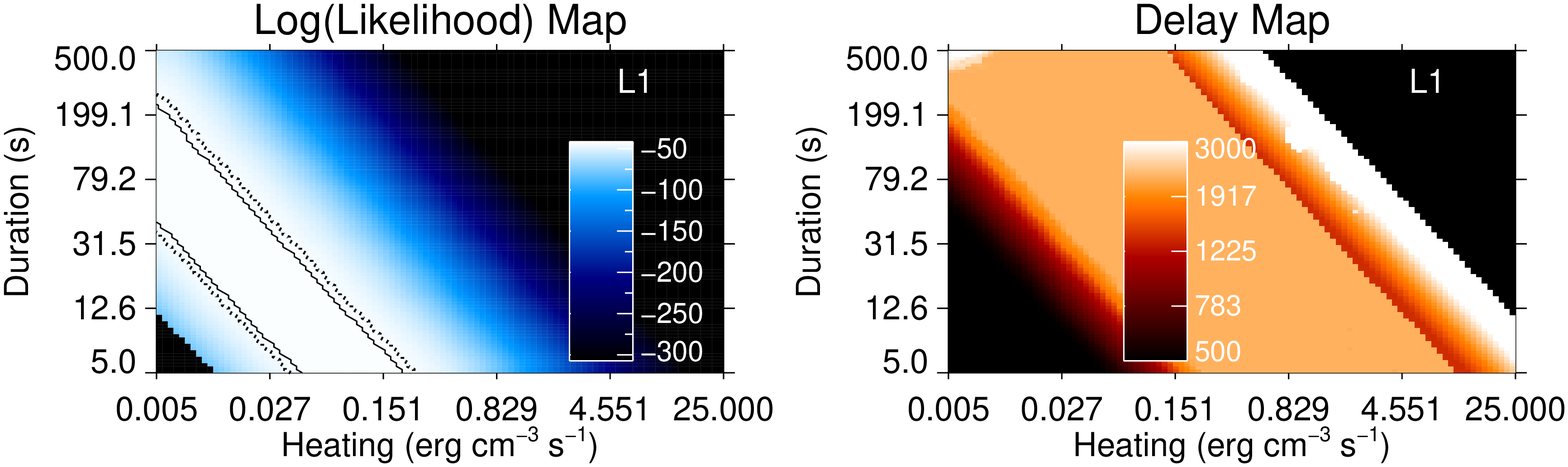} 
    \end{subfigure}
    \hfill
    \begin{subfigure}
        \centering
        \includegraphics[width=0.6\linewidth]{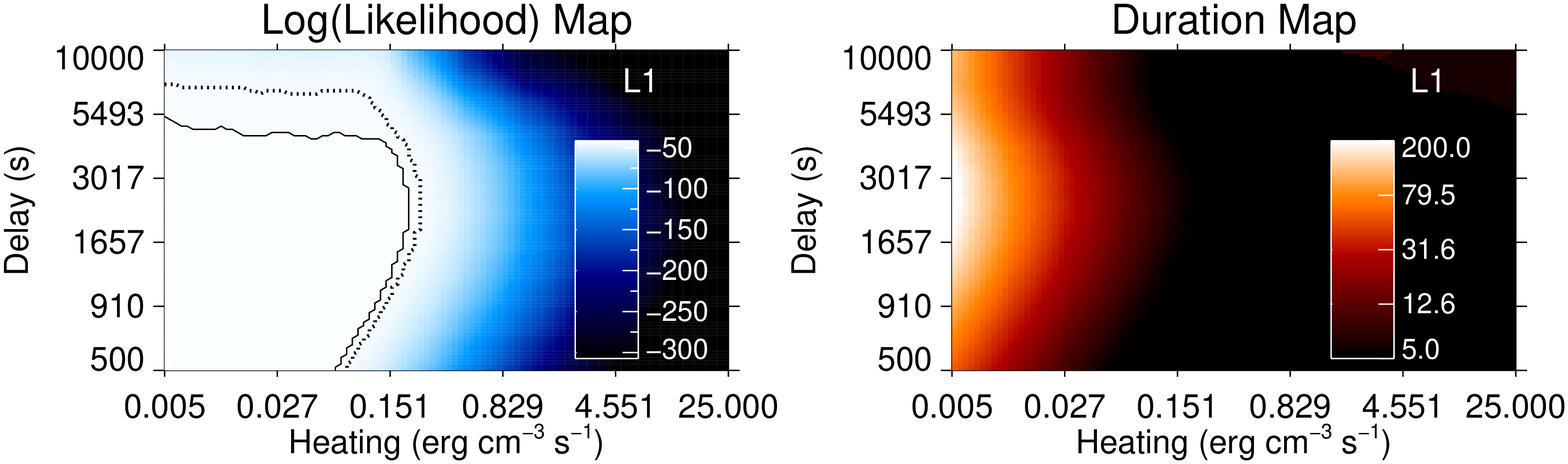} 
    \end{subfigure}
    \begin{subfigure}
        \centering
        \includegraphics[width=0.6\linewidth]{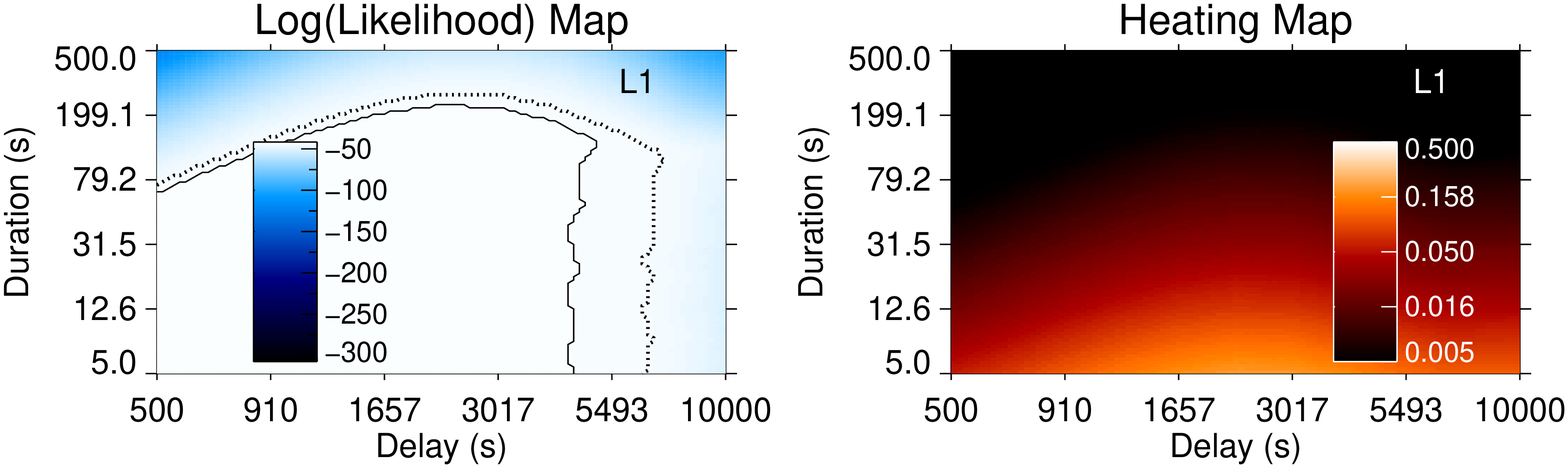} 
    \end{subfigure}
\caption{Parameter space results for two {\nustar}-observed active regions (D1 and L1) using combined data from both telescopes (FPMA \& FPMB). (Left) 2D log likelihood intensity maps for each combination of $H_0$, $\tau$, and $t_N$. (Right) Intensity maps of the optimized third parameter corresponding to each 2D likelihood plot. Energy flux constraints (Equation \ref{eqn:flux_limit}) and EUV/SXR limits from AIA and XRT have been applied to the full parameter space. The likelihood maps were smoothed for display purposes using a Gaussian kernel of width $\sigma$=1 pixel. Solid lines in the left panels show 90\% CIs and dotted lines show 99\% CIs for the case of 3 relevant parameters.}
\label{fig:nustar_D1_L1}
\end{figure*}

\begin{figure}
\centering
  \begin{minipage}[c]{\columnwidth}
\includegraphics[width=\columnwidth]{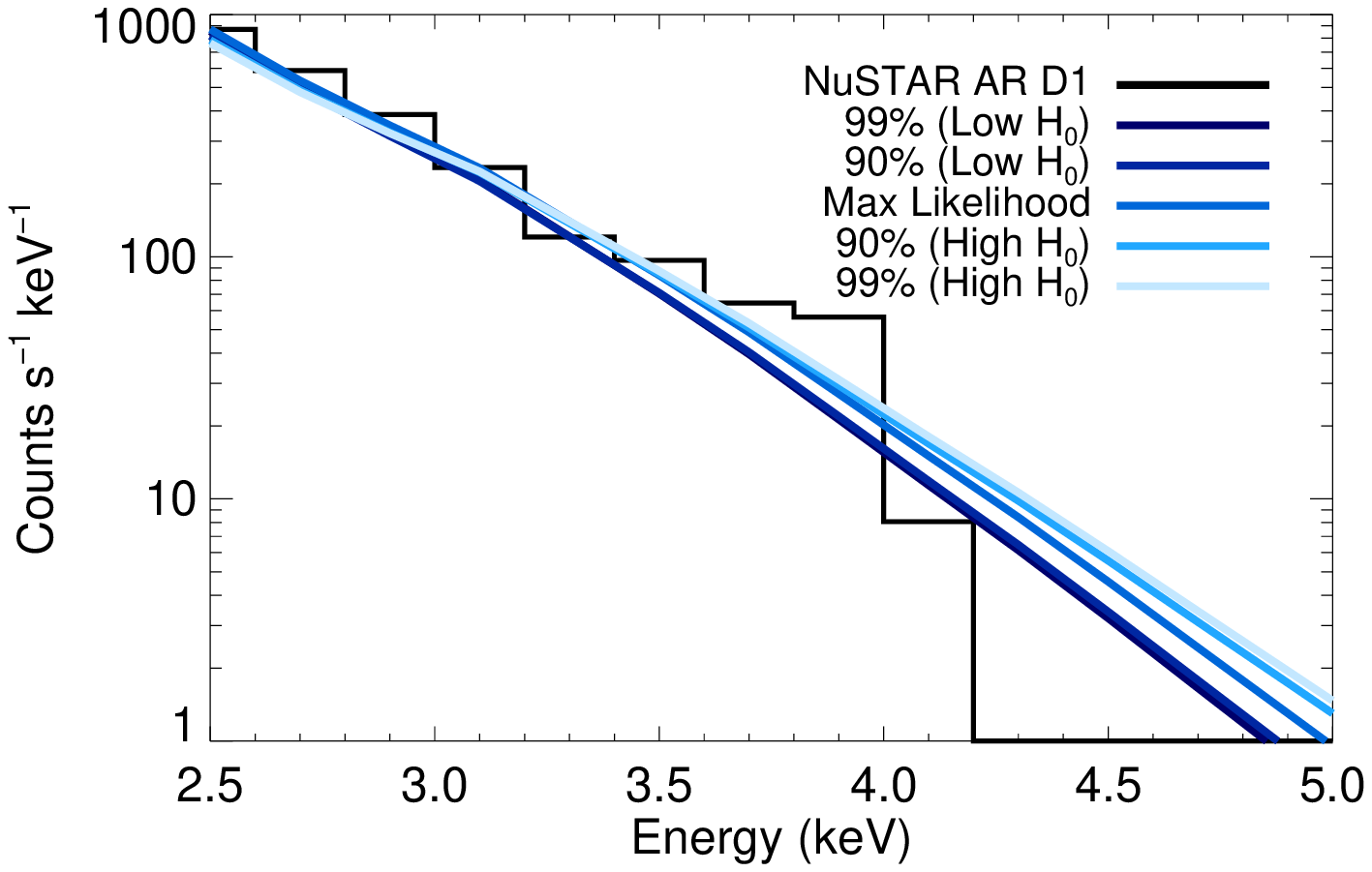}
  \end{minipage}\hfill
  \begin{minipage}[c]{\columnwidth}
  	\footnotesize
	\begin{tabular}{|c|c|c|c|c|}
	\hline
					&	$H_0$	&	$\tau$ &	$t_N$	&	$f$	\\
					&	(erg~cm$^{-3}$~s$^{-1}$) & (s) & (s) & \\
	\hline
	99\% (Low $H_0$)	&	0.025	&	12.6	&	500		&	0.29	\\
	90\% (Low $H_0$)	&	0.027	&	12.6	&	500		&	0.22	\\
	Max Likelihood		&	0.039	&	12.6	&	617		&	0.13	\\
	90\% (High $H_0$)	&	0.13		&	12.6	&	2237		&	0.093  \\
	99\% (High $H_0$)	&	0.14		&	12.6	&	2237		&	0.069  \\
	\hline
	\end{tabular}
  \end{minipage}\hfill
\caption{{\nustar} FPMA count spectrum of AR D1 and and simulated FPMA spectra at five points in the optimized, constrained heating vs. duration parameter space (Figure \ref{fig:nustar_D1_L1}). For a fixed duration $\tau$ = 12.6 s, we chose heating amplitudes at the maximum likelihood as well as on the 90\% and 99\% contours at lower and higher heating values.  The heating parameters corresponding to each curve are specified in the table. }
\label{fig:nustar_multimodel}
\end{figure}

At the 99\% confidence level there is only a 1\% a priori probability that the parameters of interest fall outside the corresponding CIs; therefore we used this confidence level to estimate the acceptable parameter ranges for each active region.  From the upper left panel of Figure \ref{fig:foxsi_all_limits}, we can see that heating amplitudes between 0.02 and 13 erg cm$^{-2}$ s$^{-1}$ are required for good agreement with the \textit{FOXSI-2} count spectra. The nanoflare duration and delay are essentially unconstrained for this AR, although delays $<$ 900 s result in slightly poorer fits and are excluded by the 90\% CIs. Steady heating (the top left corner of the delay vs. duration plot) is ruled out by the 99\% CI. The delays in the best-fit regions of parameter space for this region, while unconstrained at long values, are consistent with previous studies of simulated emission measure distributions \citep{Car2014}, observations of transient Fe XVIII brightenings \citep{Uga2014}, and time-lag studies \citep{Via2017}. The exclusion of steady heating models is also consistent with these and other studies.  

Figure \ref{fig:foxsi_hist} shows histograms of the filling factor, normalized for each model, with no limits, energy flux limits, and AIA/XRT limits applied. Without any constraints, there is a wide range of allowed filling factors due to the normalization procedure described in Section \ref{methods}. When energy and observational constraints are applied the range of acceptable filling factors is significantly reduced; most importantly, non-physical values of $f$ $>>$ 1 are eliminated. Large (unphysical) filling factors result in extremely large DEMs and predicted fluxes at EUV/SXR wavelengths, and are therefore ruled out by AIA/XRT constraints. Extremely small filling factors are ruled out by the energy flux constraint because the parameter combinations which require tiny normalizations are nanoflare sequences with extremely large energy fluxes. While $f$ is difficult to constrain observationally, the range of allowed filling factors for nanoflare models of this active region (10$^{-7}$--1) is reasonable.   

\begin{figure*}[htp]
\centering
\includegraphics[width=0.4\textwidth]{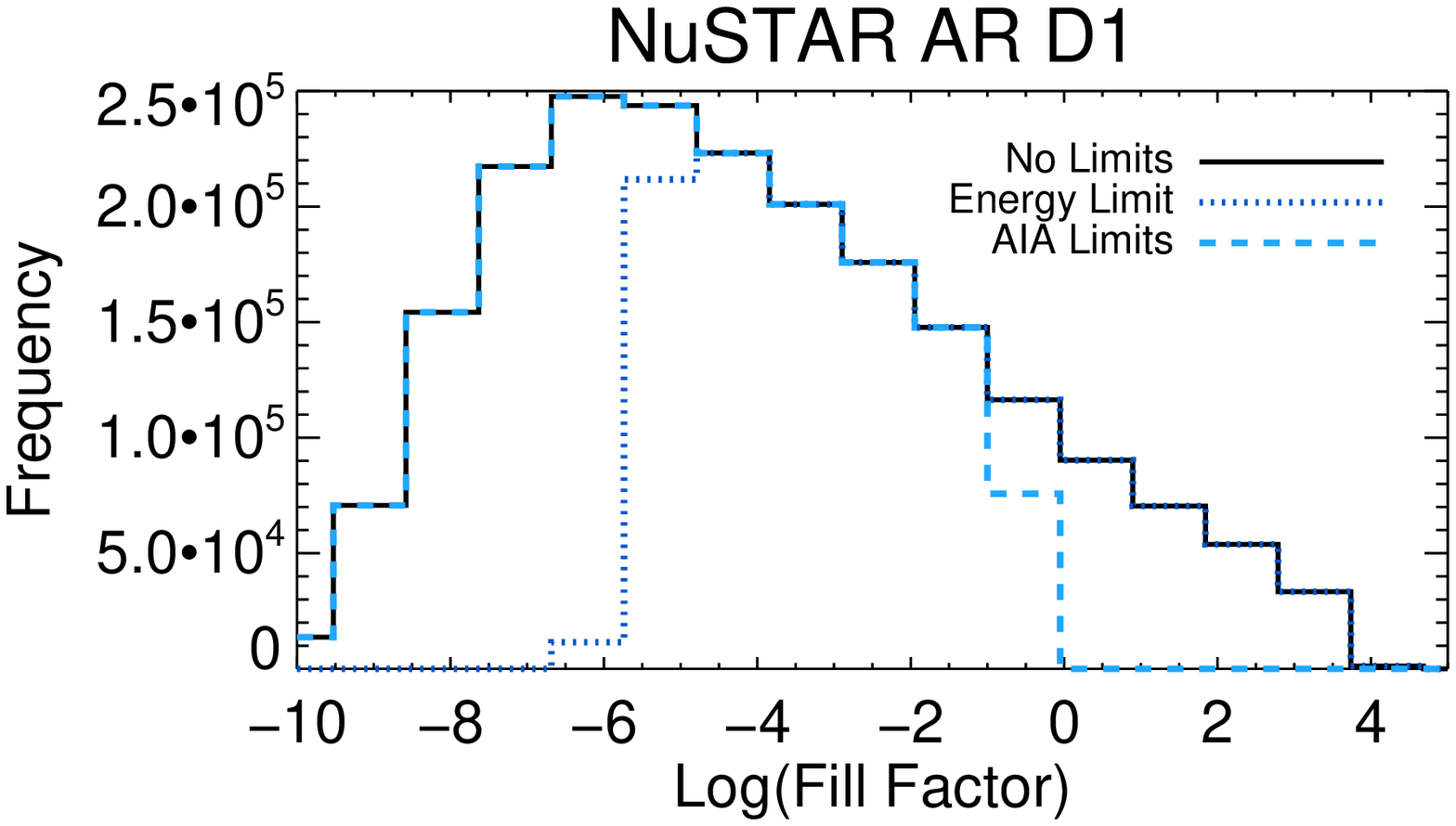}
\includegraphics[width=0.4\textwidth]{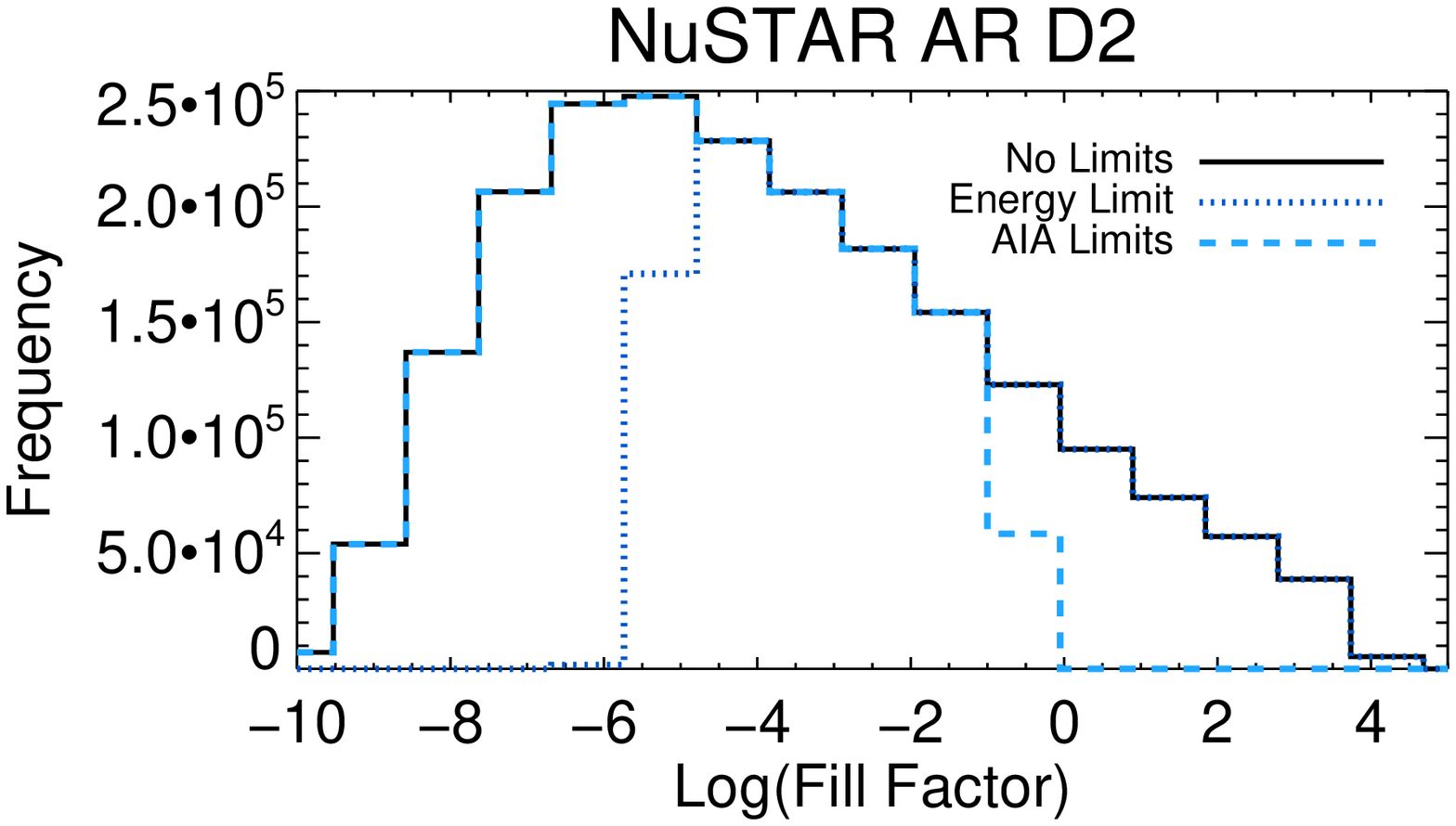}
\includegraphics[width=0.4\textwidth]{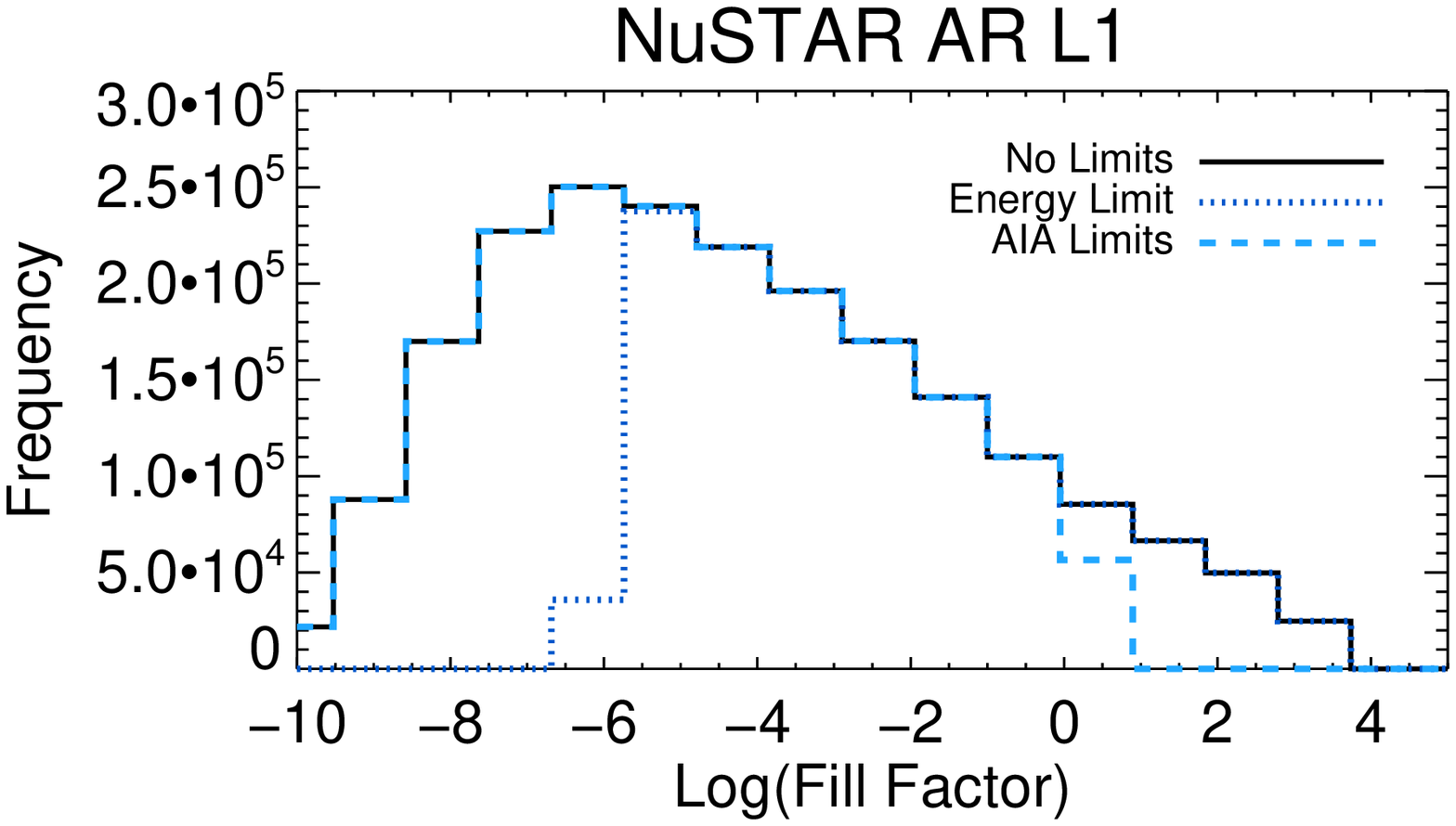}
\includegraphics[width=0.4\textwidth]{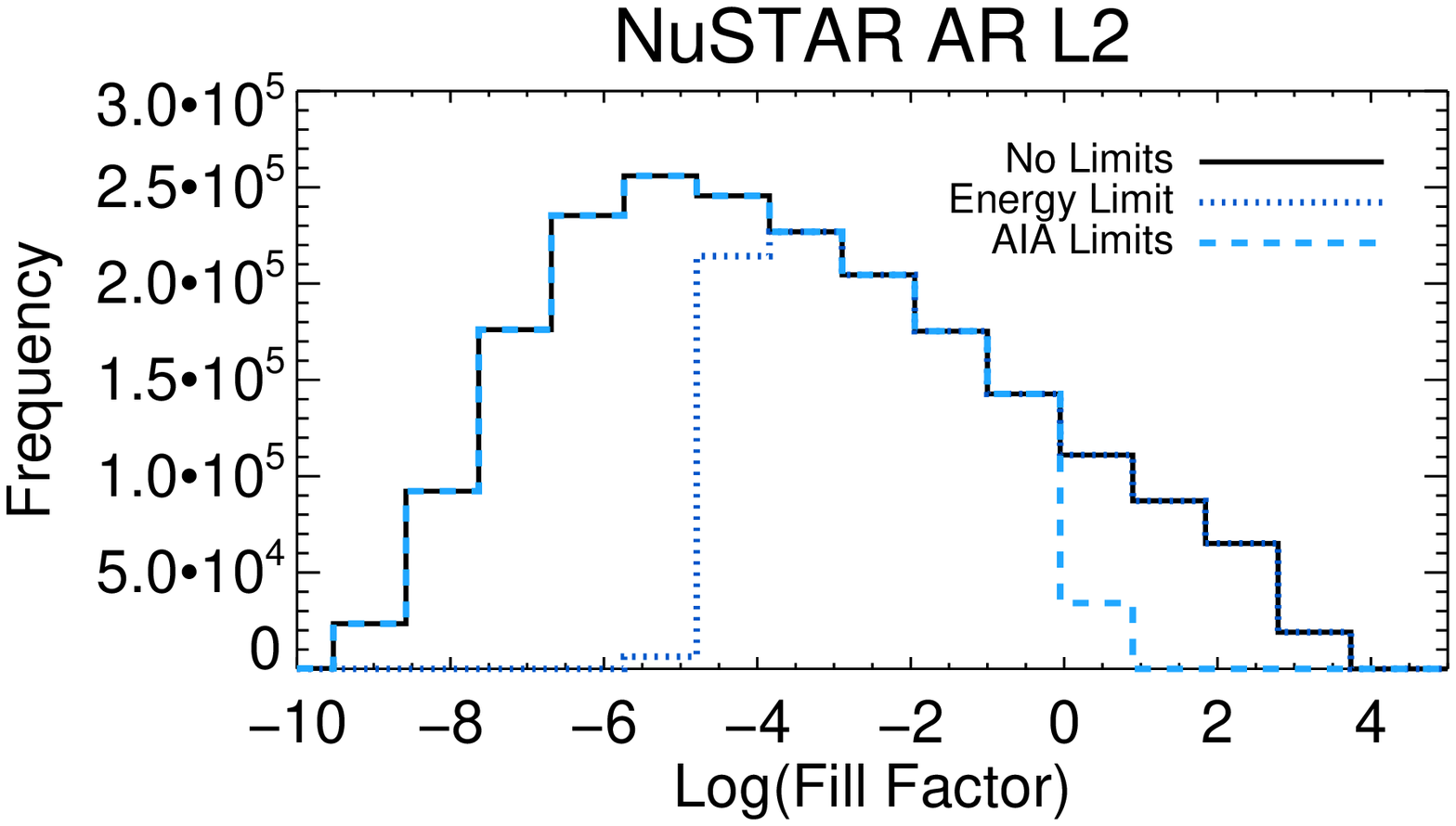}
\includegraphics[width=0.4\textwidth]{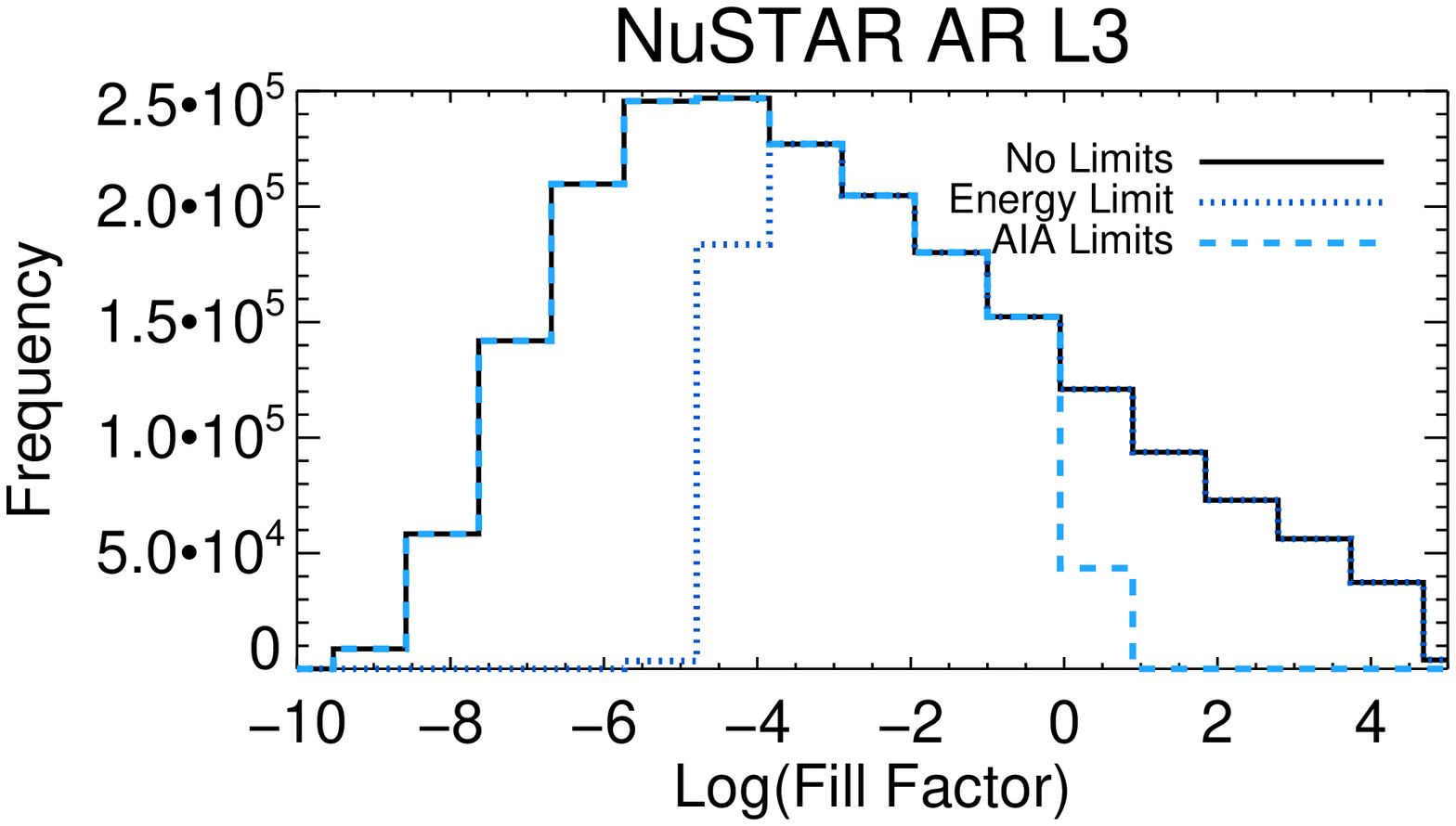}
\caption{Histograms of the fill factor for the 5 {\nustar}-observed ARs and three different sets of constraints: no limits, energy flux limits, and AIA limits.}
\label{fig:nustar_hist}
\end{figure*}

\subsection{\textit{NuSTAR} regions AR D1, L1}

Figure \ref{fig:nustar_D1_L1} shows log likelihood intensity maps and the corresponding optimized parameter maps for two of the {\nustar}-observed active regions (D1, L1), using data from both telescopes and with energy and EUV/SXR constraints imposed. Unlike the \textit{FOXSI-2} results, the {\nustar} likelihood maps were smoothed \textit{after} optimization using a Gaussian kernel of width $\sigma$=1 pixel (the parameter maps are unsmoothed). Once again, the black regions of parameter space in the two upper left panels are regions where energy flux and AIA/XRT constraints eliminated every parameter combination. The shapes of the confidence contours are noticeably different for these regions than for AR 12234. In addition, the absolute likelihoods for the {\nustar} ARs are smaller than the \textit{FOXSI-2} likelihoods due to higher counts fluxes and more data points. However, this does not mean the {\nustar} fits are poorer quality, just that the data are more constraining.    

Figure \ref{fig:nustar_multimodel} shows the {\nustar} FPMA spectrum of D1 compared to models drawn from the heating/duration 2D parameter space, similar to Figure \ref{fig:foxsi_multimodel}. The parameters for these sampled models are shown in the table below the spectrum.  

We used the 99\% CI curves to determine ranges of $H_0$, $\tau$ and $t_N$ for ARs D1 and L1. Heating amplitudes $H_0 <$ 0.32 erg cm$^{-3}$~s$^{-1}$ and $H_0 <$ 0.23 erg cm$^{-3}$~s$^{-1}$ were required for good agreement with the D1 and L1 count spectra, respectively. These maximum values are almost two orders of magnitude smaller than the maximum heating amplitude for AR 12234, which is likely due to the cooler temperatures of the {\nustar} ARs (isothermal $T$ $\sim$ 4~MK compared to $T$ $\sim$ 11~MK). Interestingly, D1 (L1) is fit well by models with $t_N$ $<$ 5000 s (7500 s), again in contrast to AR 12234 (for which the best fits occurred at $t_N$ $>$ 900 s). The duration is limited to $\tau$ $<$ 415 s for D1 and $\tau$ $<$ 275 s for L1. Even though L1 is a limb region, its likelihood and parameter maps look very similar to those of D1 and D2 (see next section). 

\subsection{\textit{NuSTAR} regions D2, L2, L3}

\begin{figure*}[htp]
\centering
    \begin{subfigure}
        \centering
        \includegraphics[width=0.6\linewidth]{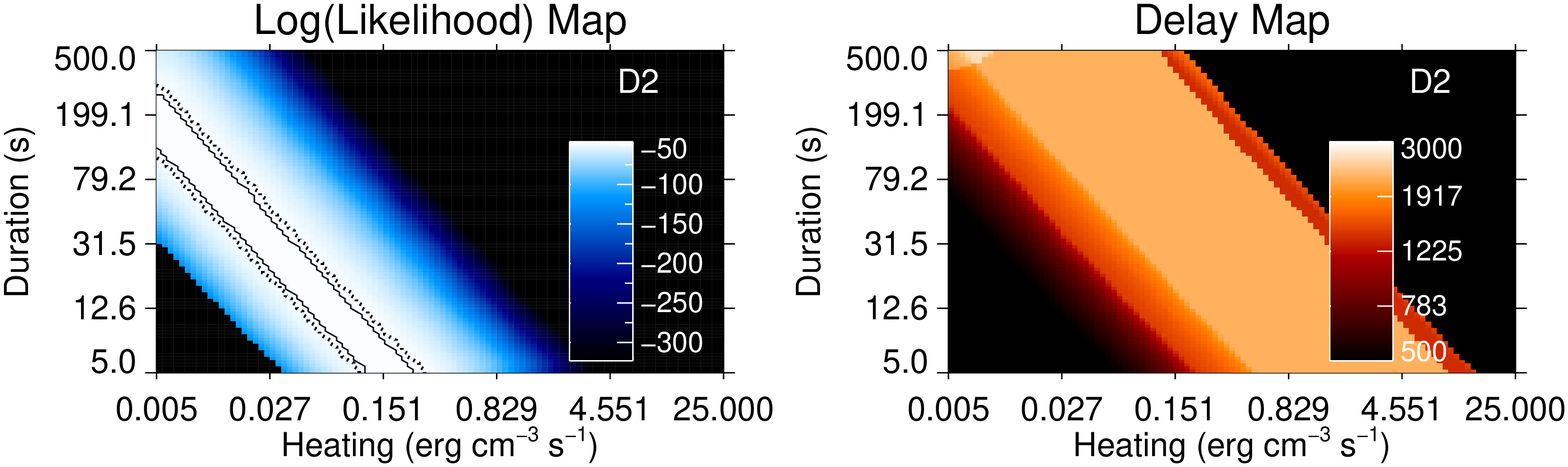} 
    \end{subfigure}
    \hfill
    \begin{subfigure}
        \centering
        \includegraphics[width=0.6\linewidth]{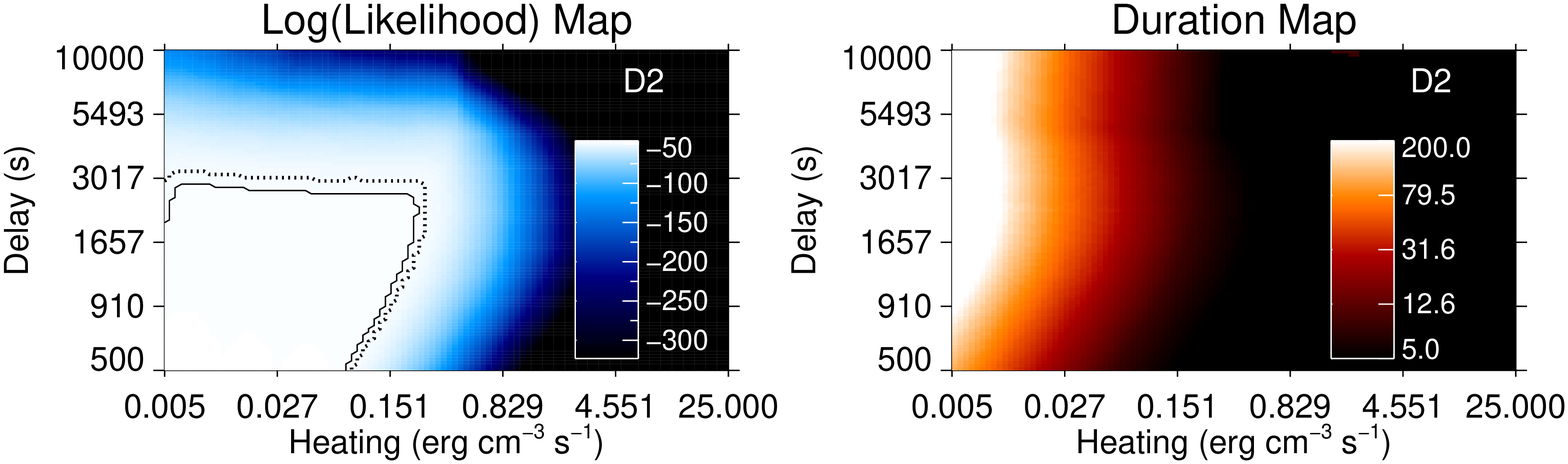}
    \end{subfigure}
    \begin{subfigure}
        \centering
        \includegraphics[width=0.6\linewidth]{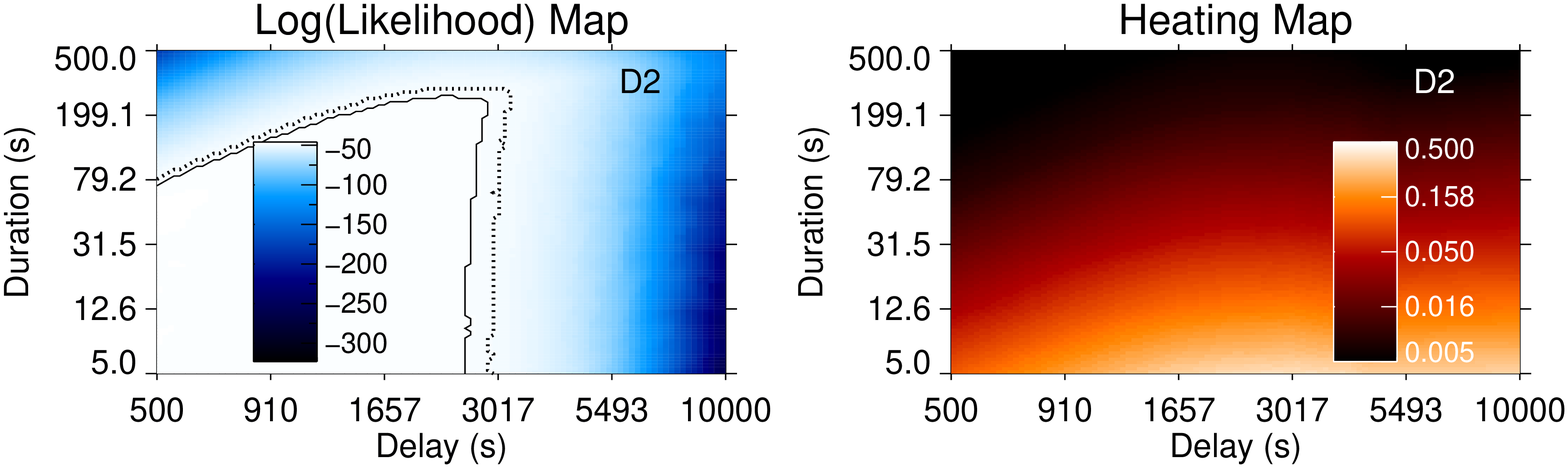} 
    \end{subfigure}
\caption{Parameter space results for {\nustar}-observed active region D2 using combined data from both telescopes (FPMA \& FPMB) and including energy flux and AIA constraints. The formatting is the same as Figure \ref{fig:nustar_D1_L1}.}
\label{fig:nustar_D2}
\end{figure*}

Figure \ref{fig:nustar_D2} shows the log likelihood intensity maps and corresponding heat maps for AR D2, an on-disk region. These maps and the CIs are fairly similar to those for D1 and L1. This is an unsurprising result because of the HXR spectral similarity between these regions, which were fit by isothermal temperatures from 3.1--4.1~MK and maximum count flux values of $\sim$10$^{3}$ counts s$^{-1}$ keV$^{-1}$ at 2.5~keV (see Figure 3 of \citealt{Han2016}). 

For this region, heating values $H_0$ $>$ 0.25 erg cm$^{-3}$ s$^{-1}$ are outside the 99\% CIs and do not yield good fits for any combination of duration and delay. Delays $t_N$ $<$ 3300 s are preferred, as are durations $\tau$ $<$ 300 s. The 99\% contours for this region are generally thinner than the same contours for ARs D1 and L1, which is most likely due to spectral differences. Separate fits to spectra from the two {\nustar} telescopes gave isothermal temperatures that differed by 0.9~MK for D2, compared to temperature differences of 0.3 and 0.2 MK for D1 and L1, respectively. The differences between these count spectra placed more stringent requirements on nanoflare models to give acceptable fits to both telescopes simultaneously. The large discrepancy for D2 was a result of its position at the edge of the {\nustar} detectors and pointing differences between FPMA and FPMB \citep{Han2016}. 

\begin{figure*}[htp]
\centering
    \begin{subfigure}
        \centering
        \includegraphics[width=0.6\linewidth]{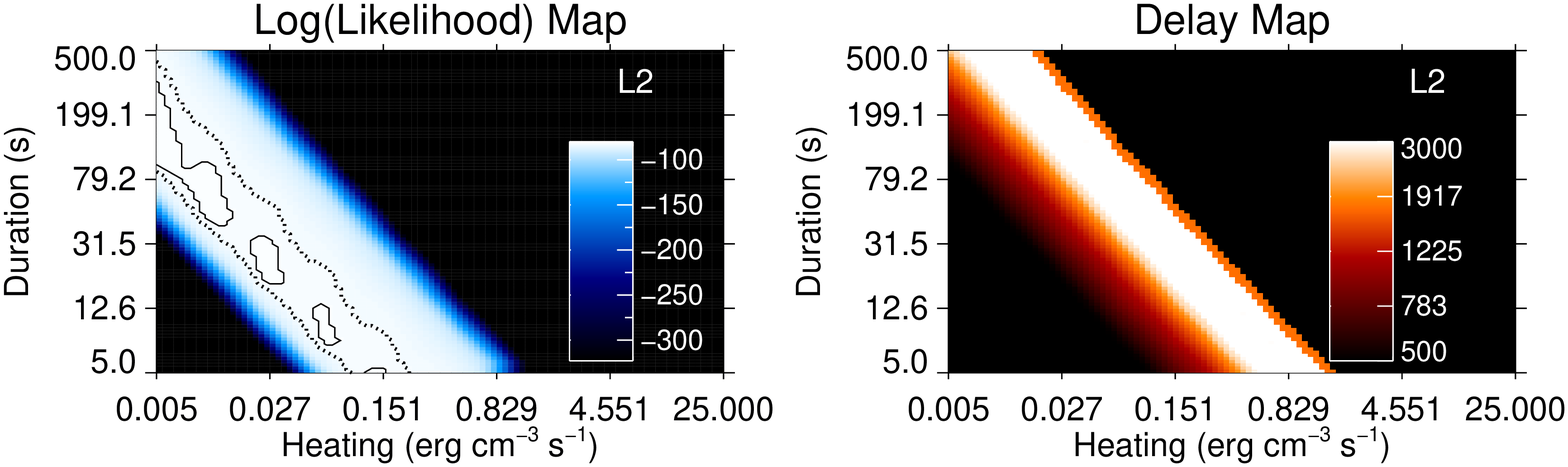} 
    \end{subfigure}
    \hfill
    \begin{subfigure}
        \centering
        \includegraphics[width=0.6\linewidth]{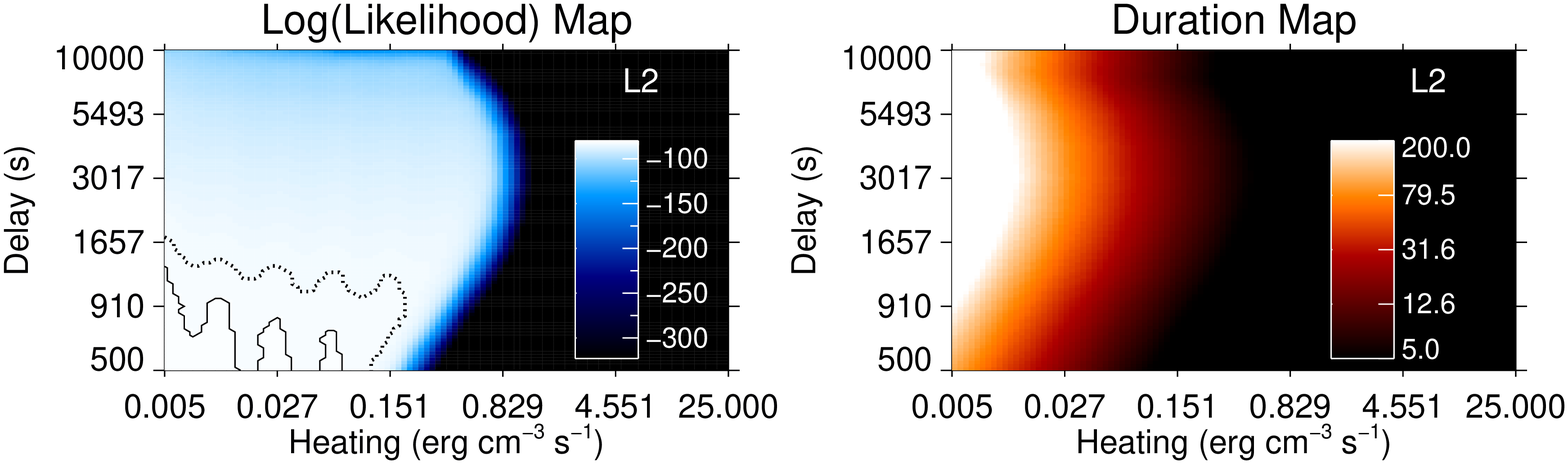}
    \end{subfigure}
    \begin{subfigure}
        \centering
        \includegraphics[width=0.6\linewidth]{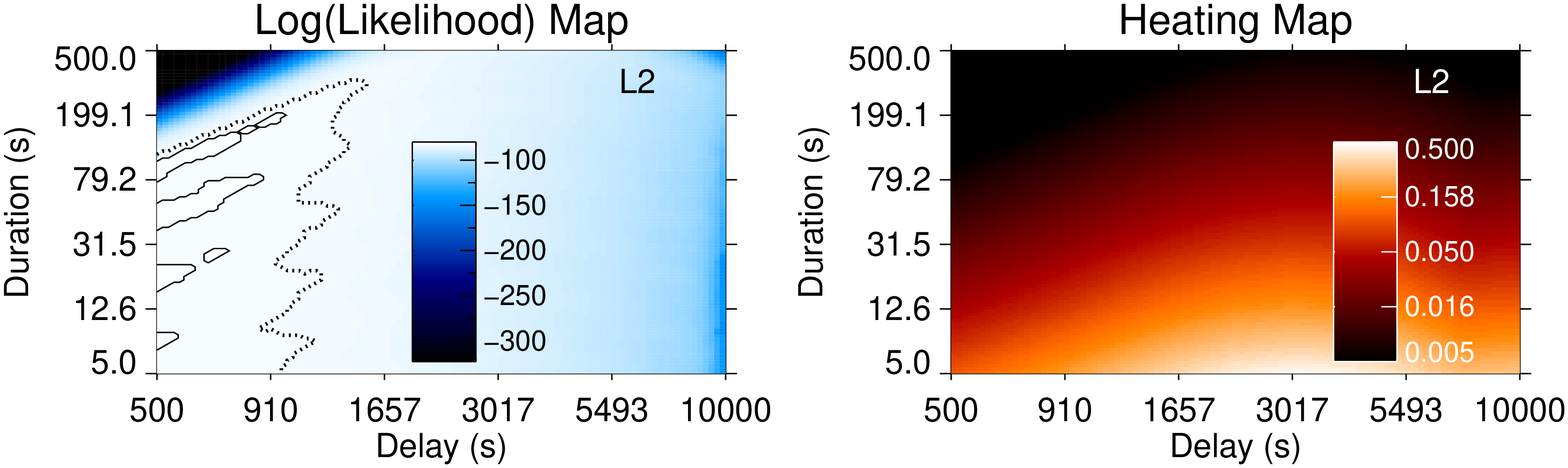} 
    \end{subfigure}
    \begin{subfigure}
        \centering
        \includegraphics[width=0.6\linewidth]{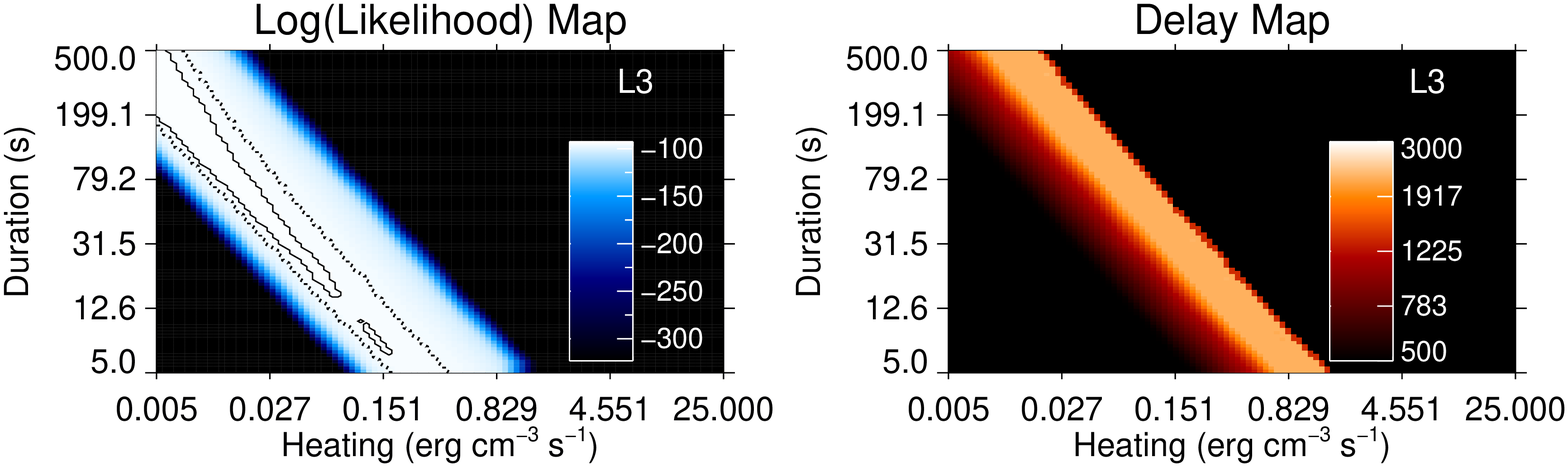} 
    \end{subfigure}
    \hfill
    \begin{subfigure}
        \centering
        \includegraphics[width=0.6\linewidth]{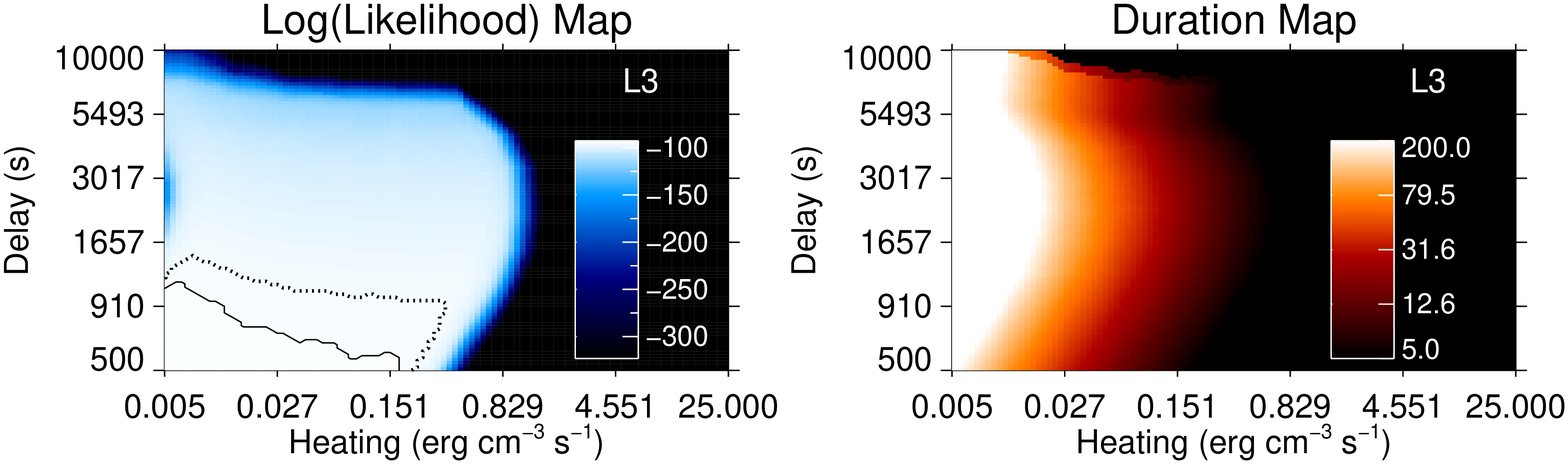} 
    \end{subfigure}
    \begin{subfigure}
        \centering
        \includegraphics[width=0.6\linewidth]{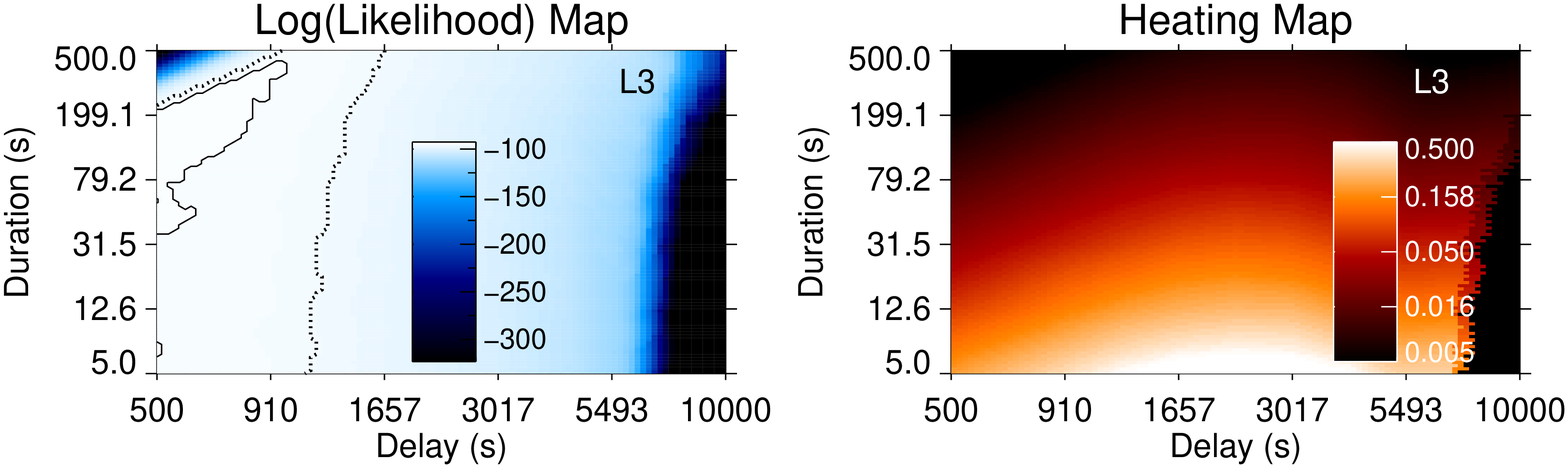} 
    \end{subfigure}
\caption{Parameter space results for {\nustar}-observed active regions L2 and L3 using combined data from both telescopes (FPMA \& FPMB) and including energy flux and AIA constraints. The formatting is the same as Figure \ref{fig:nustar_D1_L1}.}
\label{fig:nustar_L2_L3}
\end{figure*}

Figure \ref{fig:nustar_L2_L3} shows the log likelihood intensity maps and corresponding heat maps for L2 and L3, two limb regions. In contrast to the three aforementioned regions, ARs L2 and L3 were brighter and hotter (with isothermal fit temperatures between 4.1 and 4.4~MK and maximum count flux values of $\sim$10$^{4}$ counts s$^{-1}$ keV$^{-1}$ at 2.5~keV). The increased number of counts in these spectra placed stronger constraints on the model nanoflare spectra and resulted in smaller absolute likelihoods for each model (compare the likelihood colorbars from Figures \ref{fig:nustar_D2} and \ref{fig:nustar_L2_L3}). In addition, this made interpolation effects much more noticable. The gaps and other structures in Figure \ref{fig:nustar_L2_L3} are due to the interpolation of the counts flux arrays, and make it more difficult to determine accurate parameter ranges for these regions. Fortunately the 99\% CIs are fairly smooth for both these regions, and yield the following constraints for L2 and L3: $H_0$ $<$ 0.27 erg cm$^{-3}$ s$^{-1}$ and $H_0$ $<$ 0.42 erg cm$^{-3}$ s$^{-1}$, $t_N$ $<$ 1980 s and $t_N$ $<$ 1650 s, and $\tau$ $<$ 456 s and $\tau$ unconstrained respectively. 

Figure \ref{fig:nustar_hist} shows fill factor histograms for every {\nustar} AR with no constraints, energy flux constraints, and AIA constraints (no XRT data was available for this campaign). Just as in the \textit{FOXSI-2} histograms, large (unphysical) filling factors are ruled out by observational constraints and very small filling factors are ruled out by the energy flux constraint. The allowed range of $f$ for these regions is approximately 10$^{-6}$--1, values which are all physically plausible. 

\section{Conclusions}
\label{conclusions}
We modeled homogeneous sequences of nanoflares with variable heating amplitudes, durations, delays, and filling factors and compared their synthetic spectra to HXR AR spectra from {\nustar} and \textit{FOXSI-2} observations, first presented in \citet{Han2016} and \citet{Ish2017} respectively. We were able to generate good fits for the \textit{FOXSI-2} HXR data, subject to energetic and observational constraints, using homogeneous nanoflare sequences with a wide range of durations and delays. Although $t_N$ is unconstrained at the 99\% level, the best fits occur for $t_N$ $>$ 900 s in agreement with previous AR studies that did not utilize HXR data. The heating amplitudes required to fit the \textit{FOXSI-2} data are relatively high (0.02--13 erg cm$^{-2}$ s$^{-1}$), most likely because the count spectra correspond to the high-temperature ($\sim$11~MK) tail of the AR DEM. The fit quality is relatively insensitive to the nanoflare duration, which can vary from $\tau$ $<$ 5 s to $\tau$ $>$ 500 s (beyond the range of our analysis).  

For the cooler regions (characteristic temperature 3--4~MK) observed by {\nustar}, the instrument count fluxes are higher and therefore the absolute likelihoods are smaller. However, a fairly wide range of homogeneous nanoflare models yield good fits to the data (Figure \ref{fig:nustar_multimodel}). The shapes of the likelihood CIs for the {\nustar} ARs are fairly similar to each other and set limits on $H_0$, $\tau$, and $t_N$ from above, not from below. The $H_0$ vs. $\tau$ CI contours follow an approximate power-law, just like the \textit{FOXSI-2} CI contours but for smaller values of both parameters. On the other hand, the CI contours for the other {\nustar} likelihood maps ($H_0$ vs. $t_N$, $t_N$ vs. $\tau$)  are distinctly different from the corresponding \textit{FOXSI-2} AR 12234 maps. In particular, $t_N$ is bounded from above by both the 90\% and 99\% contours, as is $\tau$. $H_0$ has a smaller maximum value for these regions than for AR 12234, as well as a minimum value that is below 0.005 erg cm$^{-3}$ s$^{-1}$ (the threshold of our analysis). 

The range of acceptable parameters for each region are consistent with the temperatures derived from isothermal fits to each region's HXR spectra, although these fits characterize only a limited portion of each region's full DEM. As mentioned above, large values of $t_N$ (low-frequency heating) will result in hotter plasma than small values (high-frequency heating). It is therefore logical that the hotter \textit{FOXSI-2} AR is fit best by nanoflare sequences with longer delays, and the cooler {\nustar} ARs are fit best by nanoflare sequences with shorter delays. Similar logic can be applied to $H_0$ and $\tau$: higher values of these parameters will produce greater energy fluxes and higher temperatures. Therefore, higher heating amplitudes and longer durations should be expected to produce the best fits to AR 12234, and in fact they do. Crucially, quasi-continuous heating is excluded with $>$99\% confidence for every active region in our sample. In other words, there is no region for which the delay and duration can have the same value (500 s) within the likelihood CIs. This is a further validation of the nanoflare model, as virtually any coronal heating mechanism should be impulsive on the spatial scale of a single loop strand \citep{Kli2006,Kli2015}. 

Because \textit{FOXSI-2} and {\nustar} have limited spectral range, it is difficult to determine if the parameter space results for each instrument are different due to intrinsic properties of the ARs, or because each instrument is sampling a different component of each region's DEM distribution. According to Figure 5 of \citet{Han2016}, the best-fit parameters for \textit{FOXSI}-observed AR 12234 (T$_{high}$ = 11.6 MK, EM = 3.0$\times$10$^{43}$ cm$^{-3}$) are right at the {\nustar} 2-sigma sensitivity limit for this sample of active regions. Therefore the NuSTAR-observed regions could have had high-temperature components in their DEM distributions with similar or lower intensities as the isothermal fit to the \textit{FOXSI}-observed AR 12234. We tested the multi-thermal nature of the \textit{FOXSI}-observed region by adding additional low-temperature components to the best-fit model. First we added a model with spectral parameters roughly centered between the fit parameters from the cooler {\nustar} regions D1, D2, and L1 (T$_{low1}$ = 3.3 MK, EM = 3.5$\times$10$^{46}$ cm$^{-3}$). Next, we tried the same procedure with spectral parameters roughly centered between the fit parameters from the hotter {\nustar} regions L2 and L3 (T$_{low2}$ = 4.4 MK, EM = 5.0$\times$10$^{46}$ cm$^{-3}$). The first 2-temperature model spectrum (T$_{high}$ plus T$_{low1}$) resulted in approximately 15\% increased flux in the lowest \textit{FOXSI-2} energy bin (4-5 keV), and neglible changes above 5 keV. However, the other 2-temperature model (T$_{high}$ plus T$_{low2}$) gave fluxes $>$6 times larger in the lowest bin and fluxes $>$2 times larger in the adjacent bin. Therefore, it is certain that AR 12234 could not be fit by a 2-temperature model in which the lower T and EM were similar to what {\nustar} observed from ARs L2 and L3. However, a 2-temperature model with low-temperature parameters similar to {\nustar}-observed regions D1/D2/L1 could agree reasonably well with the \textit{FOXSI-2} AR spectrum. 

Although we were able to obtain good agreement with HXR data from homogeneous nanoflare sequences, previous work by e.g. \citet{Ree2013} and \citet{Car2014} has shown that it is difficult to produce the range of observed AR DEM slopes with equally spaced, constant energy nanoflares. \citet{Car2014} and \citet{Car2015} showed that it is possible to reproduce a broad range of slopes with nanoflare sequences if there is a correlation between the nanoflare energy and the delay between successive events. This is a more physically motivated model, as more magnetic free energy would presumably be released by (and required for) larger events. Other authors (e.g. \citealt{Bar2016b, Bra2016, Lop2016}) have used heating amplitudes drawn from a power-law distribution instead of equal-energy nanoflares. The use of power-law distributions in energy and variable delay times is beyond the scope of this analysis, but will be explored in future work.  Future work will also include the addition of ion heating to the EBTEL simulations. In addition, comparisons with field-aligned simulations can put additional constraints on which regions of parameter space can model active region HXR fluxes within the constraints of low-temperature EUV/SXR observations. Finally, {\nustar} has observed multiple active regions since 2014 November 1, several of which were quiescent and therefore suitable for nanoflare modeling studies. Future publications will model non-homogeneous nanoflares in field-line-averaged and field-aligned using data from multiple {\nustar} and \textit{FOXSI} ARs. 

\null

This paper made use of data from the {\nustar} mission, a project led by the California Institute of Technology, managed by the Jet Propulsion Laboratory, and funded by the National Aeronautics and Space Administration. This research also made use of the {\nustar} Data Analysis Software (NUSTARDAS) jointly developed by the ASI Science Data Center (ASDC, Italy) and the California Institute of Technology (USA). We thank the {\nustar} Operations, Software and Calibration teams for support with the execution and analysis of these observations. The \textit{FOXSI-2} sounding rocket was funded by NASA LCAS grant NNX11AB75G. The \textit{FOXSI} team would like to acknowledge the contributions of each member of the \textit{FOXSI} experiment team to the project, particularly our team members at ISAS for the provision of Si and CdTe detectors and at MSFC for the fabrication of the focusing optics. The authors would like to thank P.S. Athiray for providing AIA data co-temporal with the \textit{FOXSI-2} observations. Additional thanks goes to Will Barnes for helpful discussions about ebtel++ and the physics of small heating events. AJM was supported by NASA Earth and Space Science Fellowship award NNX13AM41H. LG was supported by NSF grant AGS-1429512. SJB was supported in this effort by NSF CAREER award AGS-1450230. IGH was supported by a Royal Society University Research Fellowship. This work was supported by NASA grants NNX12AJ36G and NNX14AG07G. 

\facility{NuSTAR}.

\bibliography{references}

\clearpage

\end{document}